\documentclass[10pt]{wlscirep}
\usepackage[utf8]{inputenc}
\usepackage[T1]{fontenc}
\usepackage{amsmath,amssymb,amsfonts}
\usepackage{algorithmic}
\usepackage{algorithm}
\usepackage{array}
\usepackage{subcaption}
\usepackage{textcomp}
\usepackage{url}
\usepackage{verbatim}
\usepackage{graphicx}
\usepackage{cite}
\usepackage{xcolor}
\usepackage{caption}
\usepackage{hyperref}
\usepackage{lipsum}
\usepackage[inkscapelatex=false]{svg}
\usepackage{pdfpages}
\usepackage{siunitx}
\usepackage{float}
\usepackage{cleveref}
\usepackage{dblfloatfix}
\usepackage{placeins}
\usepackage{siunitx}
\usepackage{nameref}
\DeclareSIUnit{\sample}{sample}
\title{A Metalens-based Bicycle Safety Reflector for Autonomous Vehicle Radars}

\author[1]{Sepideh Ghasemi}
\author[1]{Jimmy Hester}
\author[1]{Aline Eid}
\affil[1]{Department of Electrical Engineering and Computer Science, University of Michigan, Ann Arbor, MI 48109, USA}

\begin{abstract}
With the rising number of interactions between autonomous or sensor-assisted vehicles---especially in poor weather conditions---come the need and opportunity for a new class of bicycle safety reflectors designed to enhance cyclist visibility to radars. To this effect, the first retrodirective planar metalens-based tag operating in the millimeter-wave automotive frequency range is proposed. The compact, lightweight ($0.61~\mathrm{g}$) design consists of two layers: a metalens layer and a patch antenna pixel layer. The metalens focuses incoming plane waves from different incidence angles onto corresponding patch antenna pixels on the second layer, which re-radiate the signal back through the metalens, enabling retrodirective operation. The proposed tag was thoroughly evaluated, demonstrating reliable detection beyond 70 m and a peak monostatic radar cross section (RCS) of $3.54~\mathrm{dBsm}$ with stable retrodirectivity over $\pm 40^\circ$, providing an average gain improvement of $7.58~\mathrm{dB}$ and an RCS enhancement of $15.16~\mathrm{dB}$ relative to a lens-less reference. A realistic deployment scenario on a metallic bicycle demonstrated up to a 110x improvement in its detectability at broadside. These results highlight the potential of the proposed passive tag to operate as a low-cost, lightweight, and easily integrable bicycle safety reflector for next-generation autonomous vehicle radar systems.

\end{abstract}
\begin{document}
\flushbottom
\maketitle
\thispagestyle{empty}
\vspace{-0.2cm}
\section*{Introduction}
On Sunday, March 18, 2018, at 9:58 p.m., an autonomously driven Uber SUV collided with a woman pushing her bicycle across a highway in Phoenix, AZ, fatally injuring her \cite{ntsb2019uber}. Despite the poorly lit conditions, the experimental vehicle---equipped with state-of-the-art light detection and ranging (LiDAR) sensors, cameras, and radars---was able to detect her but failed to classify her as either a bicycle or a pedestrian until it was too late to avoid the collision.  On Tuesday, February 6th, 2024, an autonomous Waymo drove into the path of an incoming San Francisco cyclist, leading to a collision with (fortunately) much less tragic consequences \cite{reuters_waymo_2024}. These are but two of the many events exemplifying the potentially tragic interactions between automobiles operating under varying levels of autonomy (a.k.a. \textit{driver assistance}), on the one hand, and cyclists, on the other. 
Whether it is a computer or a human driving a vehicle, the accurate sensing and classification of cyclists is essential towards enhancing their safety on public roads. While optical retro-reflectors (a.k.a. \textit{cateyes}) are now standard on bicycles, these only provide recognizable reflections when properly lit, viewed, and recognized by the vehicle or the driver. In environments featuring poor lighting or dense rain, fog, or snow, these operating conditions are not met, leading to significantly increased collision risks.

Automotive radars, which are mounted under the bumpers of most new modern vehicles, are used to provide collision avoidance assistance to drivers and cut through optically opaque weather conditions, in which all other sensors' abilities are significantly degraded. \textit{What if new vehicle perception capabilities---in the form of these automotive radars, capable of peering deep into the densest atmospheres---called for the introduction of retro-reflectors that are no longer optical but, instead, mmWave?}

To create contrast for automotive radars and enhance vehicle navigation, electromagnetic markers have been proposed in two broad categories: lane markers, placed on the road surface to delineate lanes and infrastructure~\cite{feng2018lane, douglas2023low, IEEERFID2024, JRFID, yang2025design}, and feature markers attached directly to objects facing the radar (bicycles, road signs, or roadside structures) to enhance their visibility~\cite{hallbjorner2013, braun2023harmonic}. These markers are typically desired to display a retrodirective behavior with high gain to enable reliable and long-range identification by the radars from a broad range of angles. For lane markers, Feng~\textit{et al.}~\cite{feng2018lane} proposed an asymmetric trihedral reflector; however, the design is bulky, exceeds the standard pavement-marker size~\cite{3M_Marker}, and its on-road RCS measurement includes contributions from the pavement. In~\cite{douglas2023low}, a low-profile reflector was designed for radar detection of road markings, providing a high RCS of $-13~\mathrm{dBsm}$ but with only $3.5^{\circ}$ azimuthal coverage, which is insufficient in practical applications. In~\cite{IEEERFID2024}, a retrodirective lane marker based on a Yagi--Uda Van Atta configuration achieved an RCS of $-30~\mathrm{dBsm}$ over $80^{\circ}$ azimuth with a detection range of $11~\mathrm{m}$; however, its horizontal polarization made it incompatible with the vertically polarized radars used in modern vehicles, requiring a $90^{\circ}$ rotation. Therefore, to ensure polarization matching and achieve high RCS while maintaining a low profile, the authors in~\cite{JRFID} proposed a Substrate Integrated Waveguide (SIW) horn Van Atta structure implemented in a stacked architecture for seamless road integration. The design demonstrated an RCS of $-16~\mathrm{dBsm}$ over $30^\circ$ azimuth with a detection range of $25~\mathrm{m}$. While Van Atta architectures can provide the desired retrodirectivity and, when coupled with the SIW horn antennas, can offer adequate gains with a wide angular coverage, they still suffer from a major limitation: \textit{scalability}. Since the successful design of Van Atta reflectarrays relies on carefully connecting antenna pairs with transmission lines of equal electrical length around a center or axis of symmetry, scaling the structure to enhance its RCS can only asymptotically approach a maximum RCS value, as increasingly longer lines with higher transmission losses provide quickly diminishing returns. In ~\cite{yang2025design}, a stainless steel reflector covered by a wave-transparent coating for lane-marking detection was studied. However, the design is non-retrodirective and was demonstrated only over an effective detection range of approximately $10~\mathrm{m}$ which is relatively limited for a structure measuring $26\lambda \times 38\lambda$ in size. For feature markers facing the radar, a patch array Van Atta-based tag~\cite{hallbjorner2013}, and an active harmonic RFID tag~\cite{braun2023harmonic} have been explored to enhance the visibility of cyclists and roadside infrastructure. However, the patch-based Van Atta tag faces the same scalability constraint described above, while the active harmonic tag adds cost, power consumption, and the need for integration with on-board electronics. Transparent windshield-embedded reflectors were also presented for vehicle-to-vehicle visibility~\cite{vovchuk2026optically}, but they are not retrodirective and operate only at broadside incidence, can only be deployed on cars, and although their fabricated sample is $\sim 25\lambda \times 25\lambda$ in size, it would practically need to cover a full rear windshield ($\sim 260\lambda \times 260\lambda$) to achieve high RCS.

To overcome these challenges, a metalens-based marker has the potential---which remains largely unexplored---to achieve retrodirectivity through localized, passive phase control within each unit cell. In this approach, the incident power is distributed across the entire surface, with each unit cell independently contributing to the re-radiated wavefront, enabling more scalable and efficient designs at mmWave frequencies. According to Snell’s generalized law, a spatial phase gradient along a surface can control the direction of transmitted and reflected waves, enabling anomalous reflection and refraction \cite{yu2011light}---an insight that led to the development of metalenses. Metasurface lenses, or metalenses, operate as flat lenses that convert an incident plane wavefront into a spherical wavefront by imposing tailored phase profiles across the surface. This enables electromagnetic wave focusing and steering using a flat, compact, low-profile, low-loss, and reflection-less structure \cite{Grbic, pfeiffer2013millimeter} that is also easy to fabricate. In essence, metalenses manipulate wavefronts by controlling the spatial phase and transmission profiles of the metasurface \cite{Khorasaninejad2017}. Metalens designs have been demonstrated in the sub-$6~\mathrm{GHz}$ range \cite{li2020characterization}, the X-band \cite{pesarakloo2022planar,liu2020design,liu2022low,datta2022gradient,chen2024focus}, near $19~\mathrm{GHz}$ \cite{guo2018high}, and the Ka-band \cite{lee2021single}. \textit{Nevertheless, retrodirective radar markers based on planar metalenses have yet to be realized in the millimeter-wave automotive frequency range.}

In \cite{arbabi2017planar}, a two-layer optical metasurface was proposed to perform retroreflectivity over $60^\circ$ of angular coverage at optical frequencies. The first metasurface layer, performs a spatial Fourier transformation directing the light with different incident angles to different spots on the second layer which operates as a gradient metasurface that adds spatially varying momentum equal to twice that of the incident wave, but with opposite sign. In \cite{lynch2024intersection}, a retrodirective structure was introduced at mmWave frequencies using cascaded polytetrafluoroethylene (PTFE) lenses backed by an array of individual backscattering pixels. While it provides retrodirectivity in both elevation and azimuth planes, the dielectric lens is extremely bulky, and less effective for oblique incidence due to phase aberrations.

\vspace{-0.3cm}
\begin{figure}[h]
    \centering
    \includegraphics[scale=0.3]{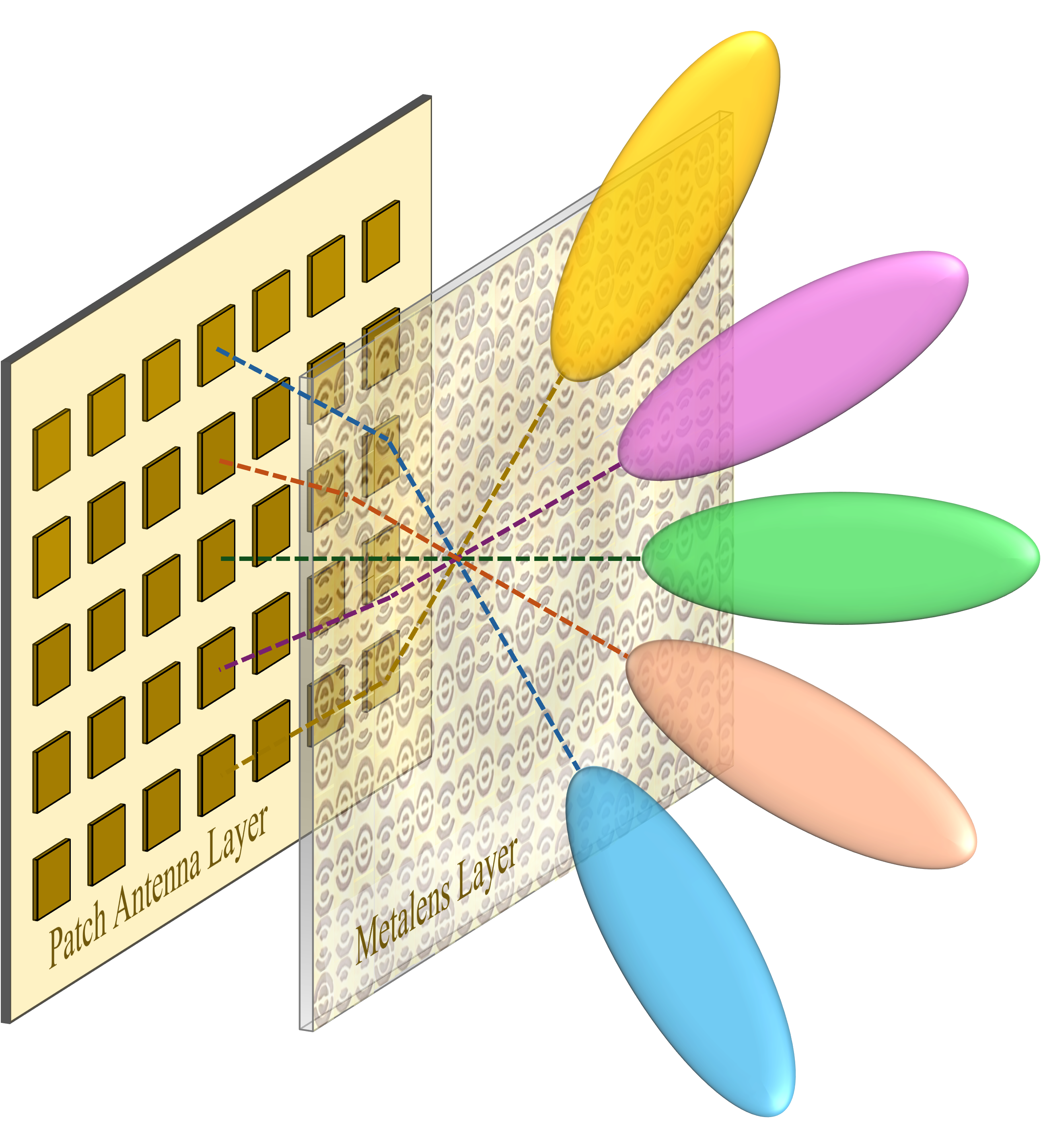}
\caption{Concept of the proposed two-layer retrodirective metalens structure.}
   \label{fig:idea}
   \vspace{-0.2cm}
\end{figure}

To bridge this gap, we propose the first retrodirective radar marker based on a planar metalens operating in the mmWave automotive radar band. The design introduces a two-layer retrodirective structure that combines and extends the advantages of \cite{arbabi2017planar} and \cite{lynch2024intersection}, as illustrated in Fig.~\ref{fig:idea}. The first layer is a metalens that focuses incoming plane waves at different oblique angles onto distinct spots on the second layer. As the waves pass through the metalens, it imparts phase shifts to direct the energy toward specific locations on the second layer, which consists of antenna pixels that receive the signal and re-radiate it back through the metalens. The reflected waves are then focused back toward the direction of incidence, achieving planar retrodirectivity. For a lens with the same aperture made of PTFE, the estimated weight would be approximately $57.95~\mathrm{g}$, whereas the proposed metalens-based tag achieves planar retrodirectivity while weighing only $0.61~\mathrm{g}$, demonstrating a lightweight and integrable mmWave solution.

\vspace{+0.1cm}

\begin{figure}[h]
\vspace{-0.15cm}
    \centering
\hspace*{1cm} 
 \begin{subfigure}{0.15\columnwidth}
\includegraphics[{width=\textwidth}]{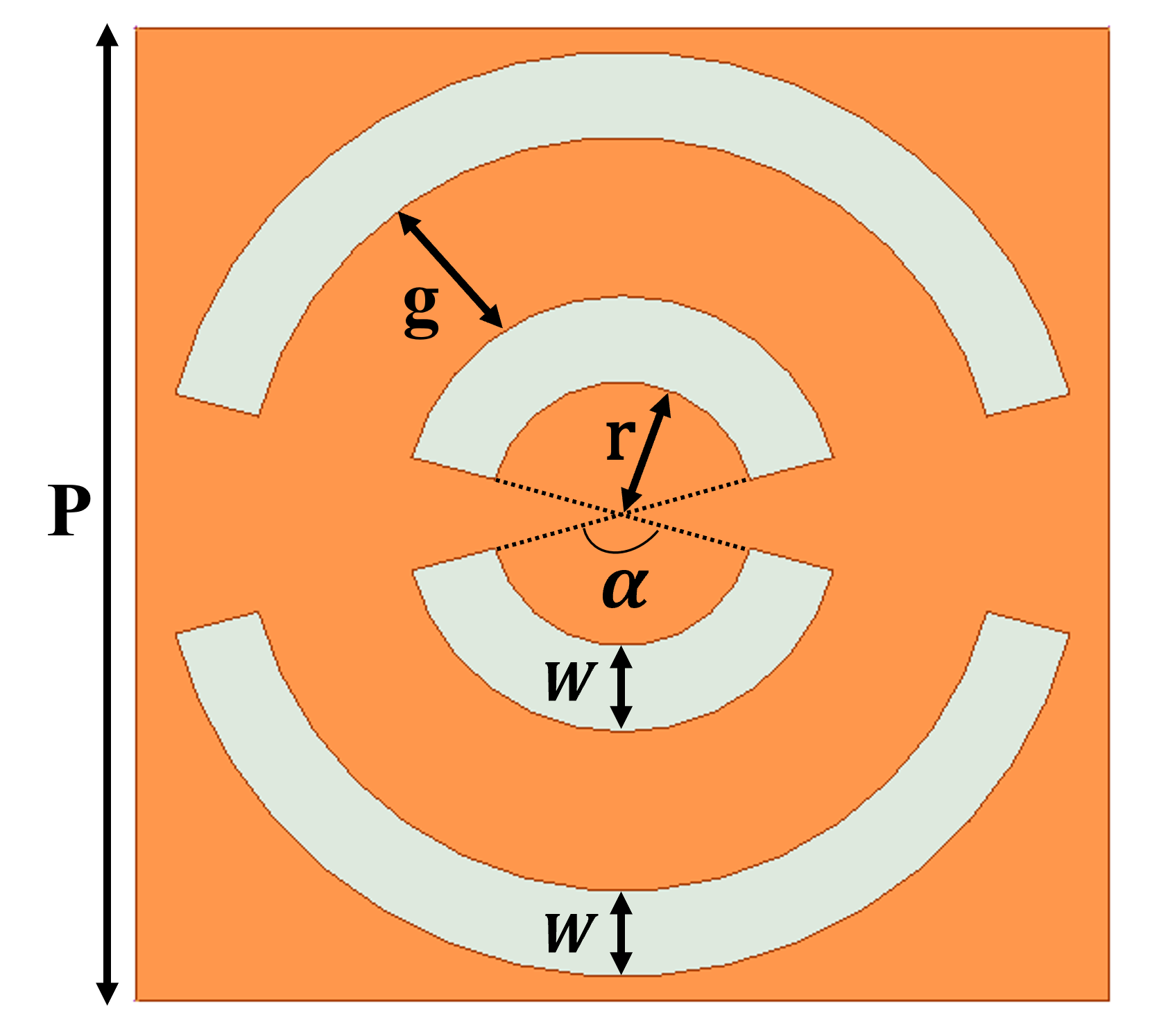}
       \caption{}
       \label{fig:2D_unitcell}
     \end{subfigure}
    \begin{subfigure}{0.15\columnwidth}
       \includegraphics[width=\textwidth]{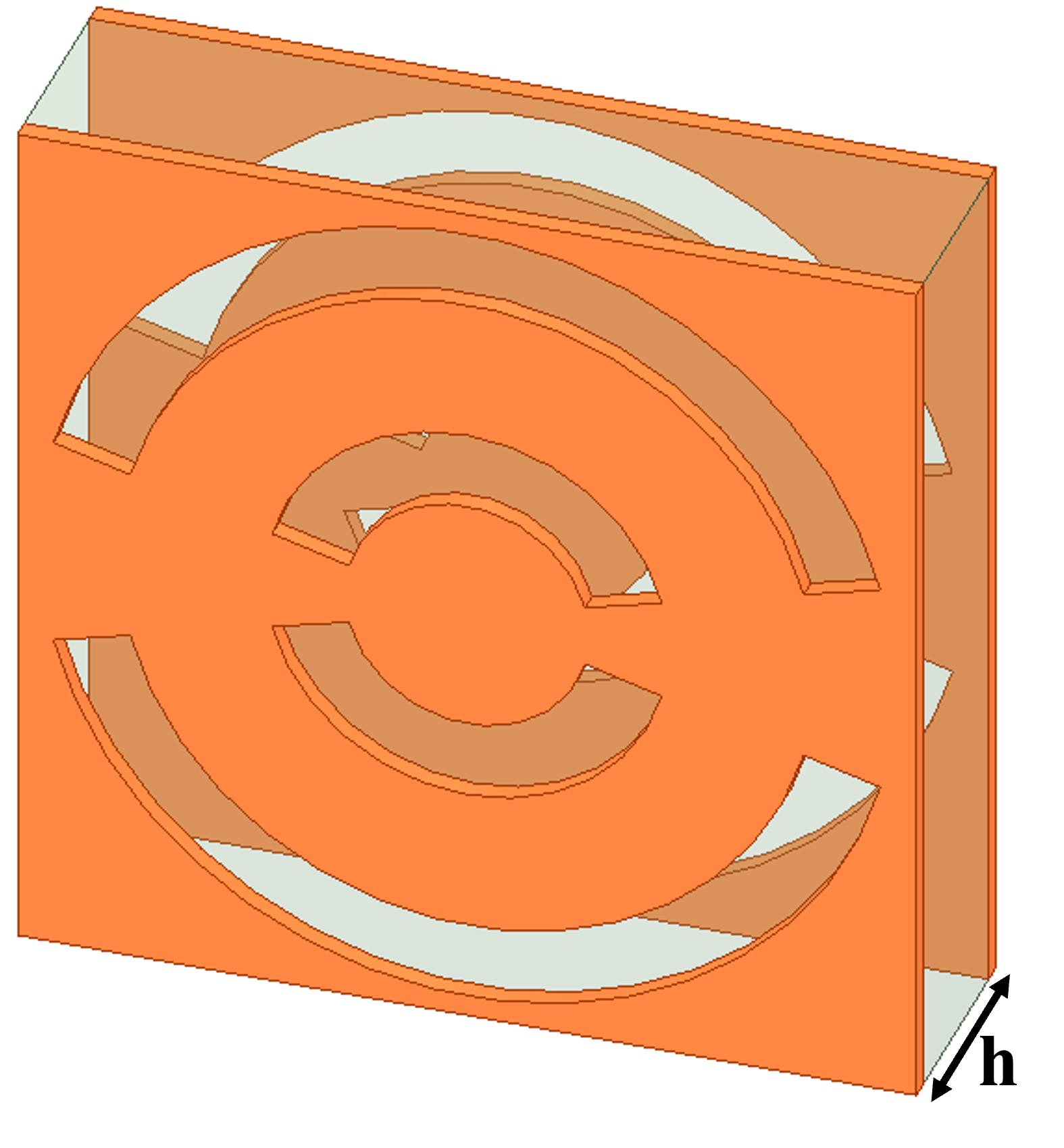}
       \caption{}
       \label{fig:3D_unitcell}
       \end{subfigure}
       \caption{(a) Top and (b) side views of the unit cell.}
	\label{fig:geometry}
    %\vspace{-0.3cm}
	\end{figure}

\section*{First Layer: Planar Metalens Design}
\subsection*{Choice of Unit Cell}
The first layer of the tag consists of a single-layer phase-gradient metalens based on a complementary double-ring resonator unit cell, designed to operate at a central frequency of 78.5\,GHz with a subwavelength thickness of $h = 0.065\lambda = 0.25~\mathrm{mm}$, as illustrated in Fig.~\ref{fig:geometry}. The structure is implemented on a single dielectric substrate (RO4350, $\varepsilon_r = 3.48$, $\tan\delta = 0.004$) sandwiched between two metallic layers patterned with complementary double rings. Each unit cell is parameterized by the radius $r$, inner and outer ring widths $w$, gap size $g$, and split angle $\alpha$, with a periodicity of $p = 1.728~\mathrm{mm}$. The top and side views of the proposed metalens are shown in Fig.~\ref{fig:2D_unitcell} and Fig.~\ref{fig:3D_unitcell}, respectively. One of the primary challenges in metalens design is achieving full $2\pi$ phase coverage while maintaining high transmission magnitude. While multilayer metasurface stacking has been proposed to overcome the limited phase range of single-layer structures while maintaining high transmission~\cite{cai2017high,cheng2019realizing,liu2022low}, such approaches increase design complexity and thickness, making them unsuitable for compact implementations. Here, rather than employing multiple substrate layers, full $2\pi$ phase coverage with high transmission is achieved through simultaneous tuning of unit-cell parameters and characterization of the corresponding transmission response. Furthermore, combining the azimuthally symmetric double-ring metalens geometry with the underlying patch layer enables efficient retrodirectivity, ensuring reliable tag operation under both normal and oblique incidence.

\vspace{+0.2cm}

\begin{figure}[h]
\centering
\begin{subfigure}{0.45\linewidth}
\centering
\includegraphics[width=\linewidth]{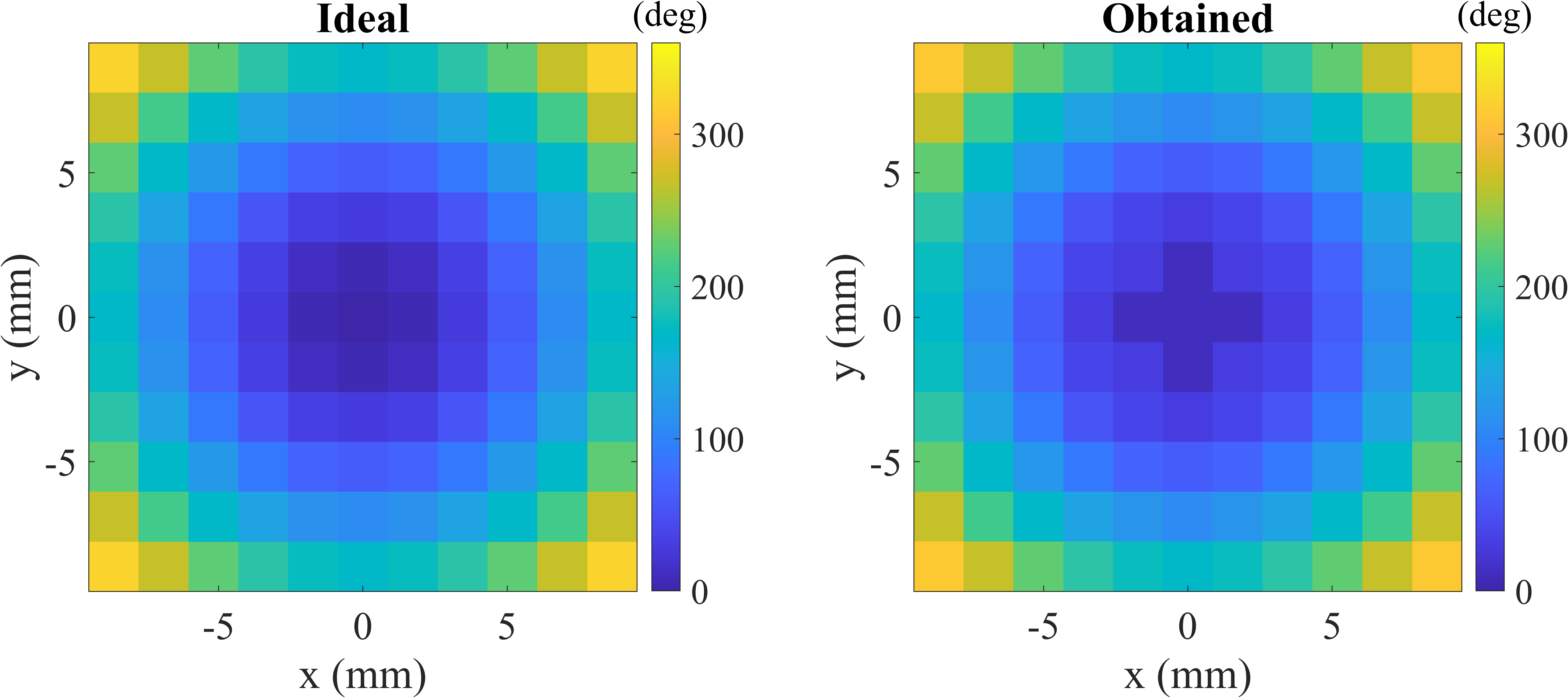}
\caption{}
\label{fig:2D_phase_profile_11unitcell}
\end{subfigure}
\hspace*{0.06\columnwidth}
\begin{subfigure}{0.45\linewidth}
\centering
\includegraphics[width=\linewidth]{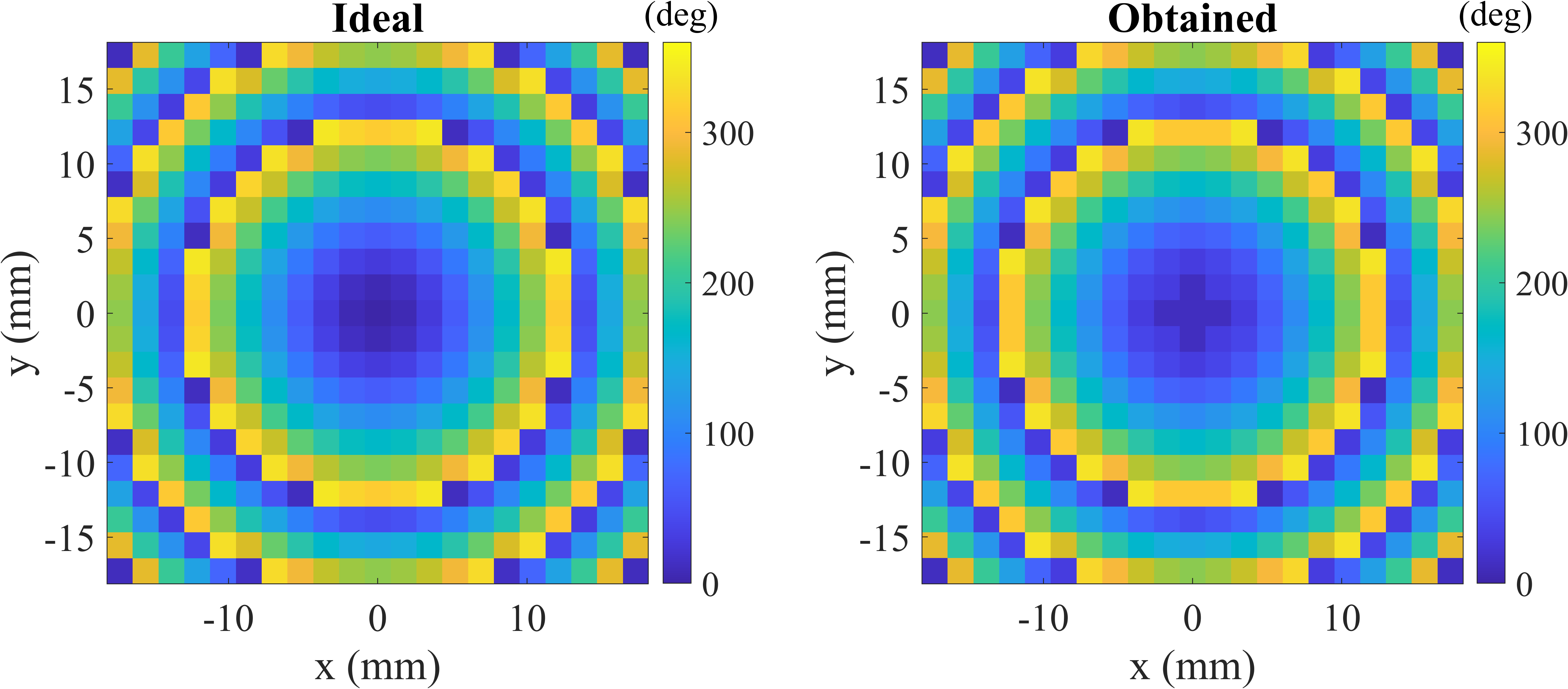}
\caption{}
\label{fig:2D_phase_profile_21unitcell}
\end{subfigure}

\caption{Required and simulated transmission phases of the 2D metalens for (a) 11 and (b) 21 unit cells.}
\label{fig:2D_phase_profile}
\end{figure}

\subsection*{Design Principles and Results} 

To focus an incoming plane wave on a focal point at a distance $f$, the phase distribution $\varphi$ across the metalens should satisfy the parabolic phase profile equation:

\begin{equation}
\varphi(x, y) = \frac{2\pi}{\lambda} \left( \sqrt{x^2 + y^2 + f^2} - f \right),
\label{eq:parabolic_phase}
\end{equation}

Here, $\lambda$ denotes the operating wavelength, with $x$ and $y$ defining the plane of the metalens, and the focal length $f$ is set to $20~\mathrm{mm}$. This parabolic phase profile ensures that the electrical length from every point on the lens to the focal point is equal, so all wavefronts constructively interfere at the focus. Based on Eq.~\ref{eq:parabolic_phase}, the required phase at each unit cell can be determined. In HFSS, the unit cell shown in Fig.~\ref{fig:3D_unitcell} was placed between two Floquet ports. By simultaneously varying the unit-cell geometry and recording the obtained transmission magnitude and phase, a library was generated. Fig.~\ref{fig:2D_phase_profile} compares the required and simulated transmission phase distributions of the 2D metalens for two cases, corresponding to 11 and 21 unit cells across the aperture. As observed in Figs.~\ref{fig:2D_phase_profile_11unitcell}--\ref{fig:2D_phase_profile_21unitcell}, a smaller number of unit cells results in rapid phase variation between adjacent elements due to coarse phase sampling. Increasing the number of unit cells produces a smoother phase gradient that more closely approximates the ideal continuous profile, thereby improving the focusing performance of the metalens. Accordingly, an aperture comprising 21 unit cells per side was selected to ensure accurate focusing while maintaining structural compactness. From this HFSS-generated library, the closest matching phase response value to the required phase at each unit cell position is identified, and the corresponding geometry parameters are assigned to that unit cell, as shown in Table~\ref{tab:map}. The phase mapping begins at the central unit cell of the 2D metalens $(0,0)$ and proceeds upward to $(0,10)$. Owing to the azimuthal symmetry of the phase distribution, the remaining unit cells of the structure can be generated based on this table. By applying this procedure across all unit cells, the corresponding metalens design is realized and fabricated, as shown in Fig.~\ref{fig:front}.

\begin{table}[!t]
\centering
\begin{tabular}{|c|c|c|c|c|c|c|}
\hline
Unit cell & $r$ ($\mathrm{mm}$) & $w$ ($\mathrm{mm}$) & $g$ ($\mathrm{mm}$) & $\alpha$ (deg) & Phase (deg) & Magnitude (linear)\\
\hline
(0,10) & 0.2 & 0.2 & 0.16 & 115 & 243.44 & 0.89\\
\hline
(0,9)  &  0.14 & 0.18 & 0.14 & 155 & 142.41 & 0.96\\
\hline
(0,8)  & 0.12 & 0.14 & 0.24 & 175 & 45.18 & 0.89\\
\hline
(0,7)  & 0.22 & 0.18 & 0.18 & 105 & 313.86 & 0.73\\
\hline
(0,6)  & 0.22 & 0.20 & 0.14 & 115 & 238.93 & 0.89\\
\hline
(0,5)  & 0.12 & 0.18 & 0.10 & 175 & 166.71 & 0.96\\
\hline
(0,4)  & 0.14 & 0.14 & 0.24 & 145 & 107.79 & 0.92\\
\hline
(0,3)  & 0.16 & 0.12 & 0.24 & 155 & 61.94 & 0.85\\
\hline
(0,2)  & 0.18 & 0.10 & 0.22 & 175 & 30.26 & 0.81\\
\hline
(0,1)  & 0.16 & 0.12 & 0.22 & 175 & 11.28 & 0.65\\
\hline
(0,0)  & 0.16 & 0.12 & 0.22 & 175 & 11.28 & 0.65\\
\hline
\end{tabular}

\caption{\label{tab:map}Mapping between required phase responses and corresponding unit-cell parameters. The parameters $r$, $g$, and $w$ are given in millimeters, and the phase and $\alpha$ are given in degrees.}
%\vspace{-0.1cm}
\end{table}

\vspace{+0.1cm}

\begin{figure}[h]
    \centering
    \hspace*{+0.5cm}
\includegraphics[{width=0.38\textwidth}]{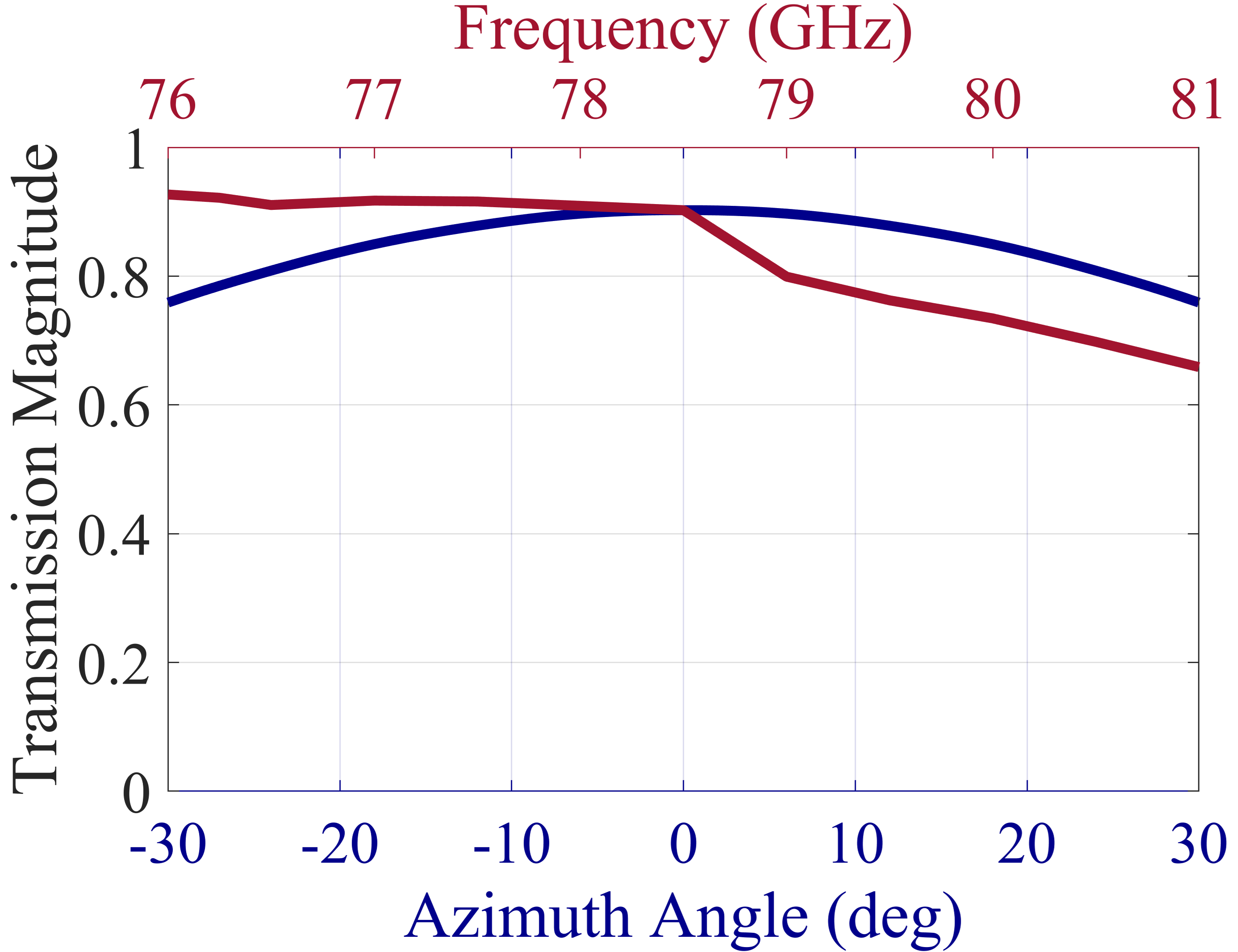}
\caption{Simulated transmission magnitude of the 2D metalens as a function of azimuth angle at $78.5~\mathrm{GHz}$ (blue) and as a function of frequency at normal incidence (red).}
\label{fig:Transmission_Angular_Frequency}
\vspace{-0.3cm}
\end{figure}

\begin{figure}[h]
\vspace{-0.4cm}
    \centering
    \hspace*{0.5cm}
\includegraphics[{width=0.38\textwidth}]{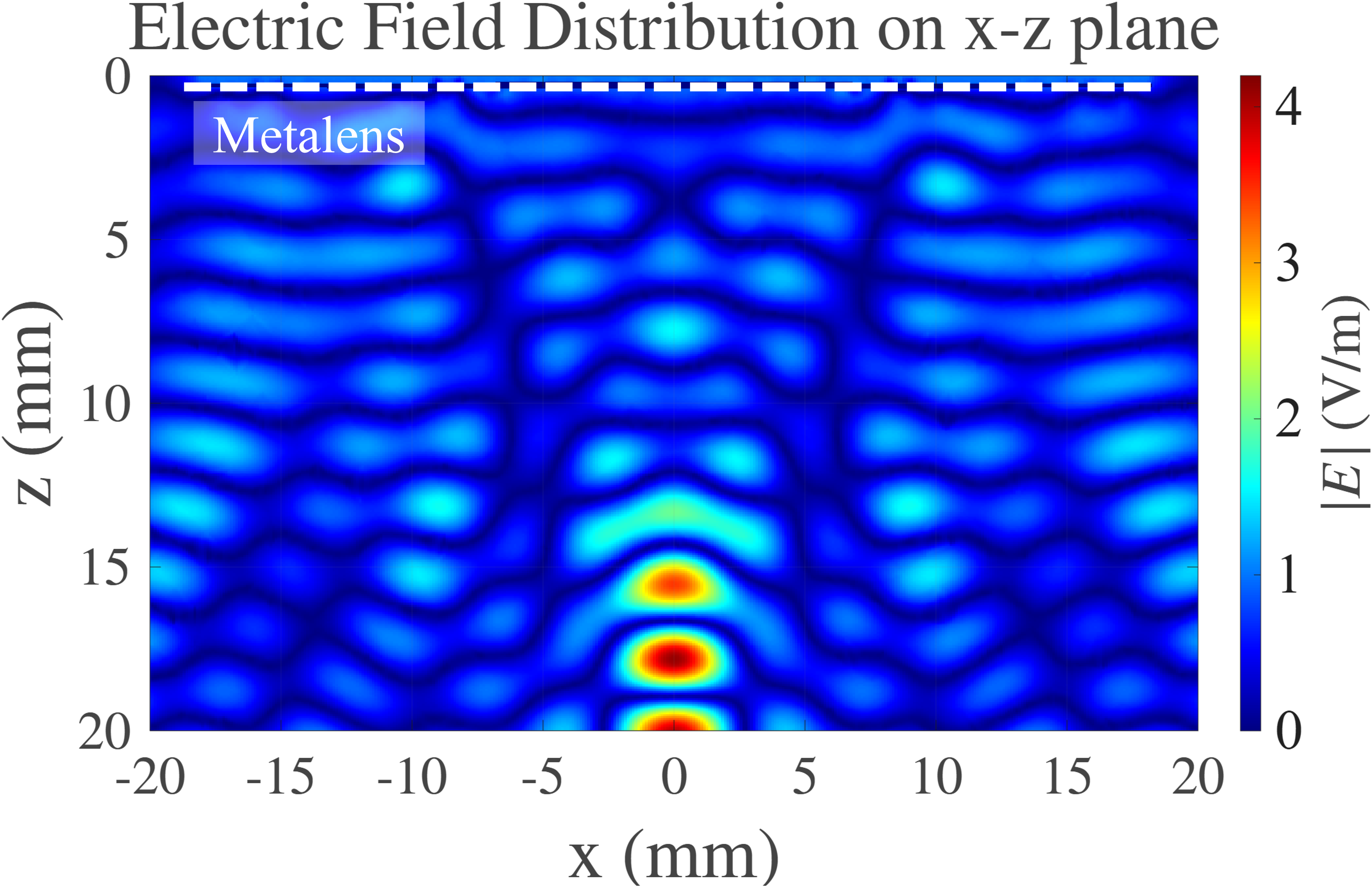}
\caption{Simulated electric-field distribution on the x-z plane, showing focusing near the designed focal point at $20~\mathrm{mm}$.}
\label{fig:1D_Metasurface_Focus}
%\vspace{-0.3cm}
\end{figure}
    
To extract the transmission magnitude of the 2D metalens under plane-wave excitation in HFSS, the electric and magnetic fields are first calculated on the incident and transmitted surfaces, each placed at a distance of $\lambda/2$ from the unit cell. The Poynting vector is then computed and integrated over both surfaces to obtain the incident and transmitted powers. The transmission coefficient is subsequently determined by normalizing the transmitted power with respect to the incident power. Fig.\ref{fig:Transmission_Angular_Frequency} shows the transmission magnitude of 2D metalens as a function of azimuth angle at $78.5~\mathrm{GHz}$ (blue) and as a function of frequency at normal incidence (red), illustrating the angular and spectral response of the metalens and indicating that a high transmission magnitude is maintained across both domains.

To validate the focusing performance of the designed metalens, the metalens is placed on the $x$–$y$ plane, with a phase gradient along both the $x$- and $y$-directions, and wave propagation along the $z$-axis. Perfectly matched layers are applied as open boundaries to emulate free-space conditions. The magnitude of the simulated electric field is plotted on an $x$–$z$ observation plane on the transmission side of the metalens. As shown in Fig.~\ref{fig:1D_Metasurface_Focus}, the transmitted waves refract toward the center and focus near the designed focal point at $20~\mathrm{mm}$, confirming the expected transmission and focusing behavior.

\section*{Second Layer: Plane of Patch Antenna Pixels}
\begin{figure}[!t]
    \centering
    % ===== ROW 1: setup + 0.4 mm + 0.6 mm =====
    \begin{subfigure}{0.25\columnwidth}
        \includegraphics[width=\textwidth]{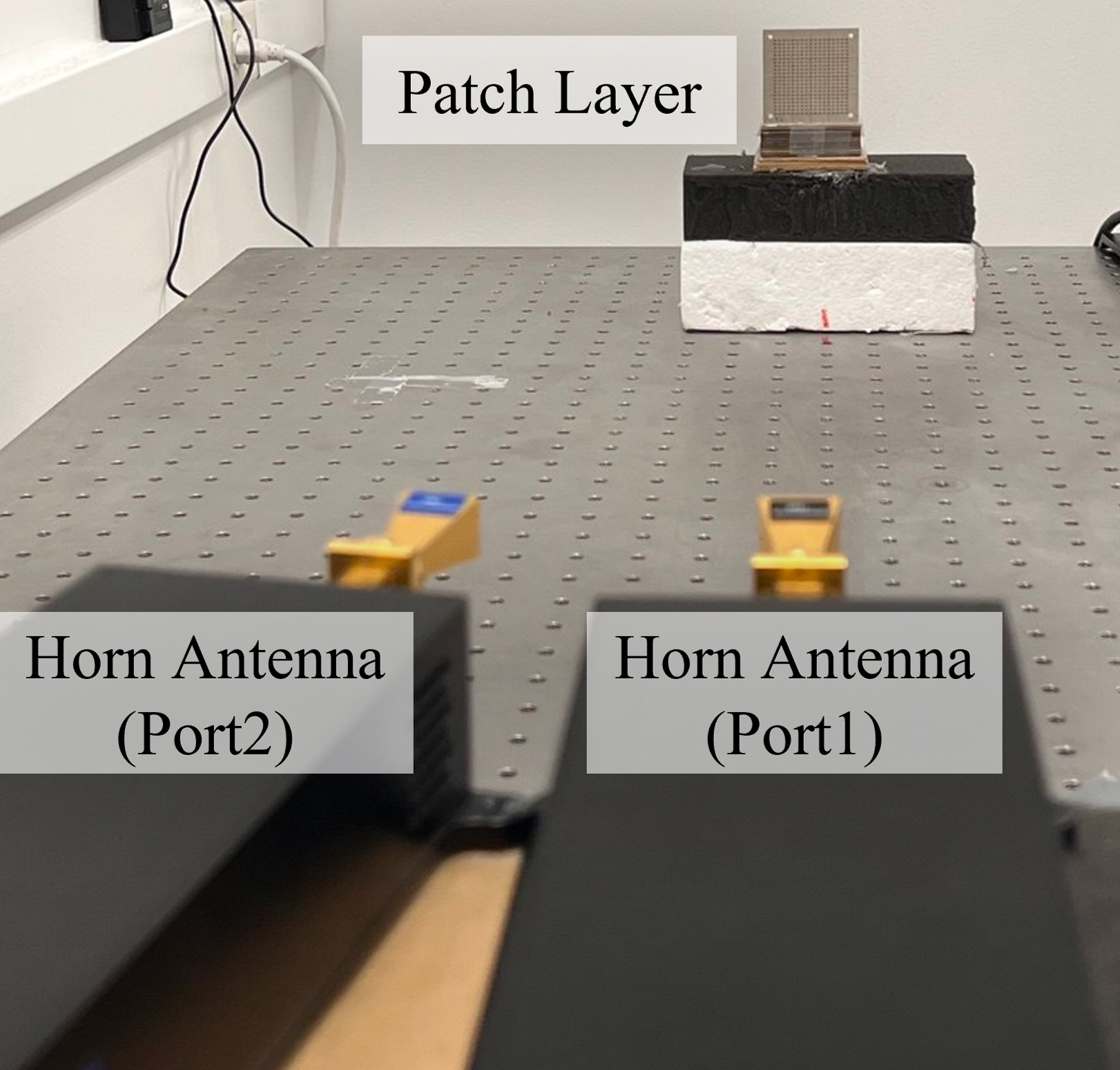}
        \caption{}
        \label{fig:patch_setup}
    \end{subfigure}
    \hspace*{0.01\columnwidth}
    \begin{subfigure}{0.38\columnwidth}
        \includegraphics[width=\textwidth]{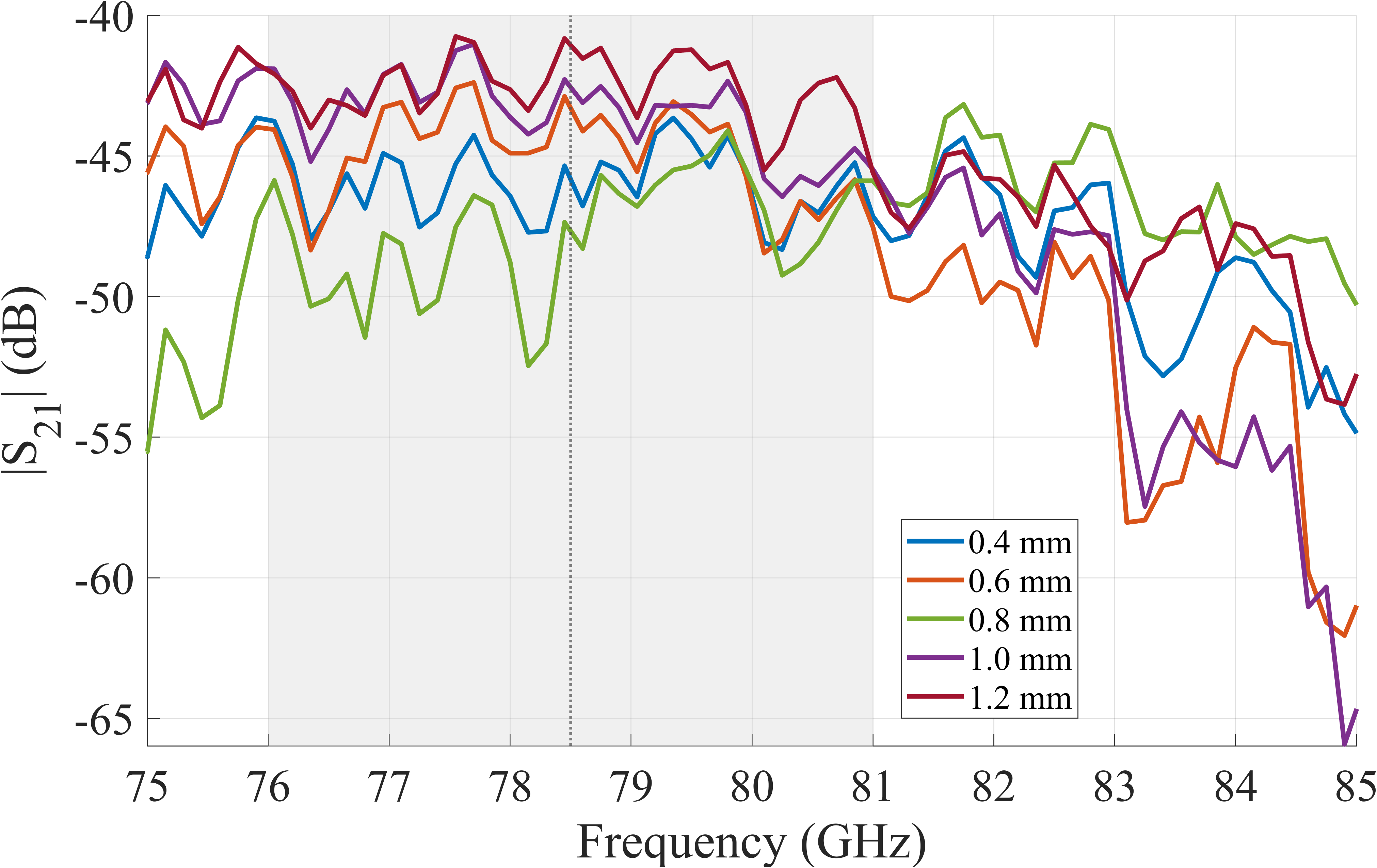}
        \caption{}
        \label{fig:Patch_S21_All_Distances}
    \end{subfigure}

    \caption{Patch-layer reflection characterization: (a) experimental measurement setup. (b) Measured transmission coefficient magnitude $|S_{21}|$ for the five fabricated patch variants with inter-patch spacings of $0.4$, $0.6$, $0.8$, $1.0$, and $1.2~\mathrm{mm}$, respectively.}
    \label{fig:patch}
    \vspace{-0.2cm}
\end{figure}

%\label{sec:Second Layer}
Following the design principles in \cite{balanis2015antenna}, the length and width of a rectangular patch antenna resonating at $78.5~\mathrm{GHz}$ were determined through HFSS simulations to be $0.84~\mathrm{mm}$ and $1.28~\mathrm{mm}$, respectively. The spacing between adjacent patches was chosen to keep the mutual coupling below $-20~\mathrm{dB}$. To verify this, two neighboring patches were simulated; when one was excited, the transmission to the other ($S_{21}$) remained below $-20~\mathrm{dB}$. Using these dimensions, the second layer of the proposed retrodirective metalens-based tag was constructed as a plane of patch-antenna pixels. To identify the optimal inter-patch spacing, five patch layers were fabricated with spacings of $0.4$, $0.6$, $0.8$, $1.0$, and $1.2~\mathrm{mm}$. For experimental characterization, each layer was mounted vertically on the optical table and positioned in front of two horn antennas connected to Port~1 and Port~2 of the VNA through millimeter-wave extenders, as shown in Fig.~\ref{fig:patch_setup}. After a two-port calibration, the horn antennas were aligned with the center of the patch layer. The horn connected to Port~1 was oriented at $0^\circ$ with respect to the patch surface; the second horn was placed at the closest achievable angle to $0^\circ$, since mechanical constraints from the extender geometry and its connection to the main measurement chain prevented both extenders from being oriented at $0^\circ$ simultaneously, thereby removing the specular reflection component. The measurement distance and orientation were kept identical across all five cases to ensure consistency. The measured transmission magnitudes $|S_{21}|$ for the five patch-layer variants are shown in Figs.~\ref{fig:Patch_S21_All_Distances}, where the maximum of $|S_{21}|$ indicates the resonance of each layer. The patch layer with $1.2~\mathrm{mm}$ inter-patch spacing exhibits the strongest resonance within the frequency band of interest ($76$--$81~\mathrm{GHz}$), with its peak at the central frequency of $78.5~\mathrm{GHz}$. This layer was therefore selected for the subsequent radar cross-section measurements.

\section*{Characterization and Implementation}
The performance of the proposed tag was evaluated using both a VNA and an automotive radar. The following subsections describe the experimental setups and corresponding results used to characterize the focal length of the metalens layer, the retrodirective behavior through monostatic RCS measurements, the detection range, and the imaging performance of the metalens-based tag.

\subsection*{Focal Length and Gain Measurement of the Metalens-based Tag}
\begin{figure}[h]
    \centering

    \begin{subfigure}{0.328\columnwidth}
        \includegraphics[width=\textwidth]{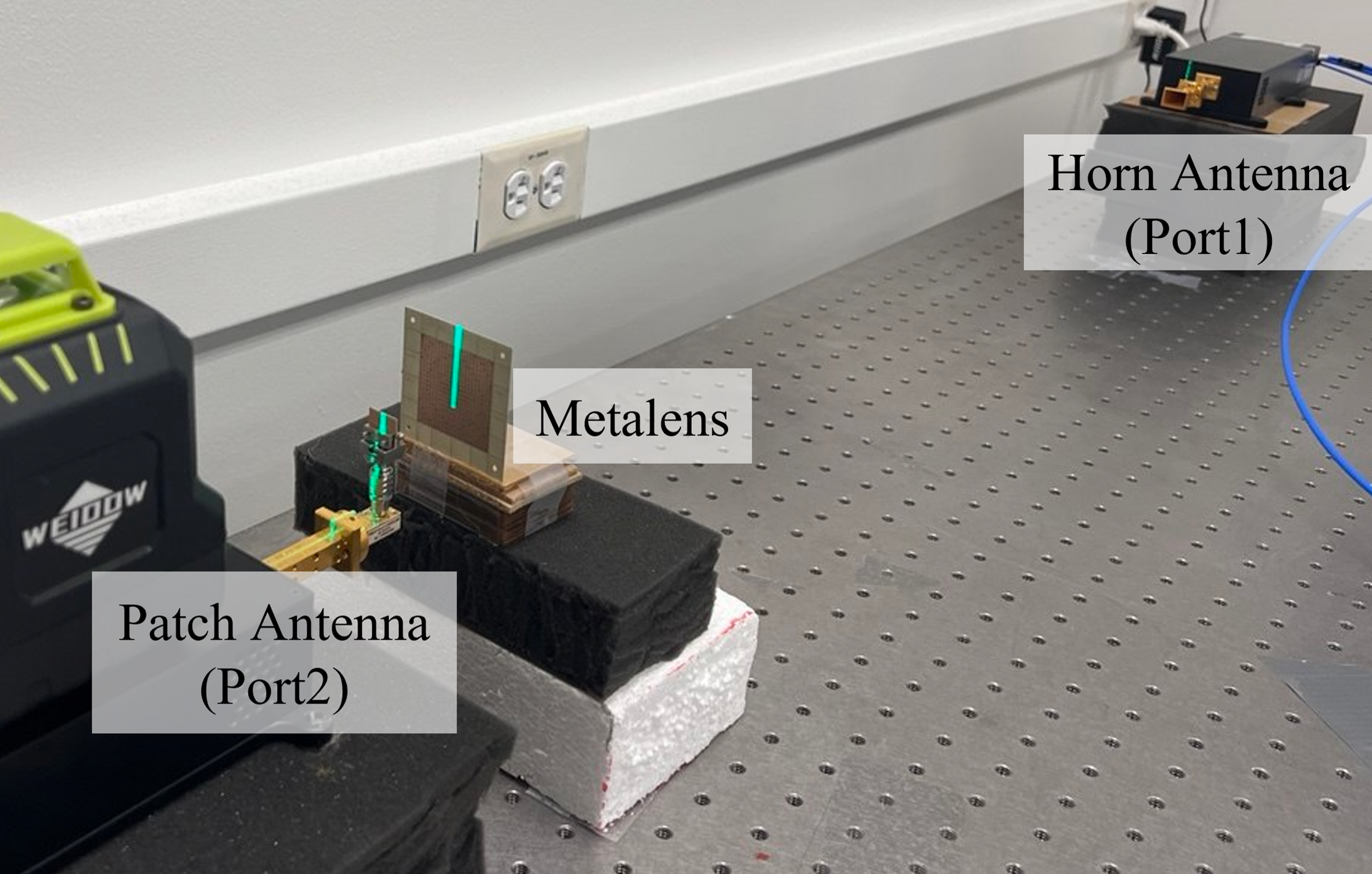}
        \caption{}
        \label{fig:Gain_Focalpoint_Setup}
    \end{subfigure}
    \hspace*{0.01\columnwidth}
    \begin{subfigure}{0.31\columnwidth}
        \includegraphics[width=\textwidth]{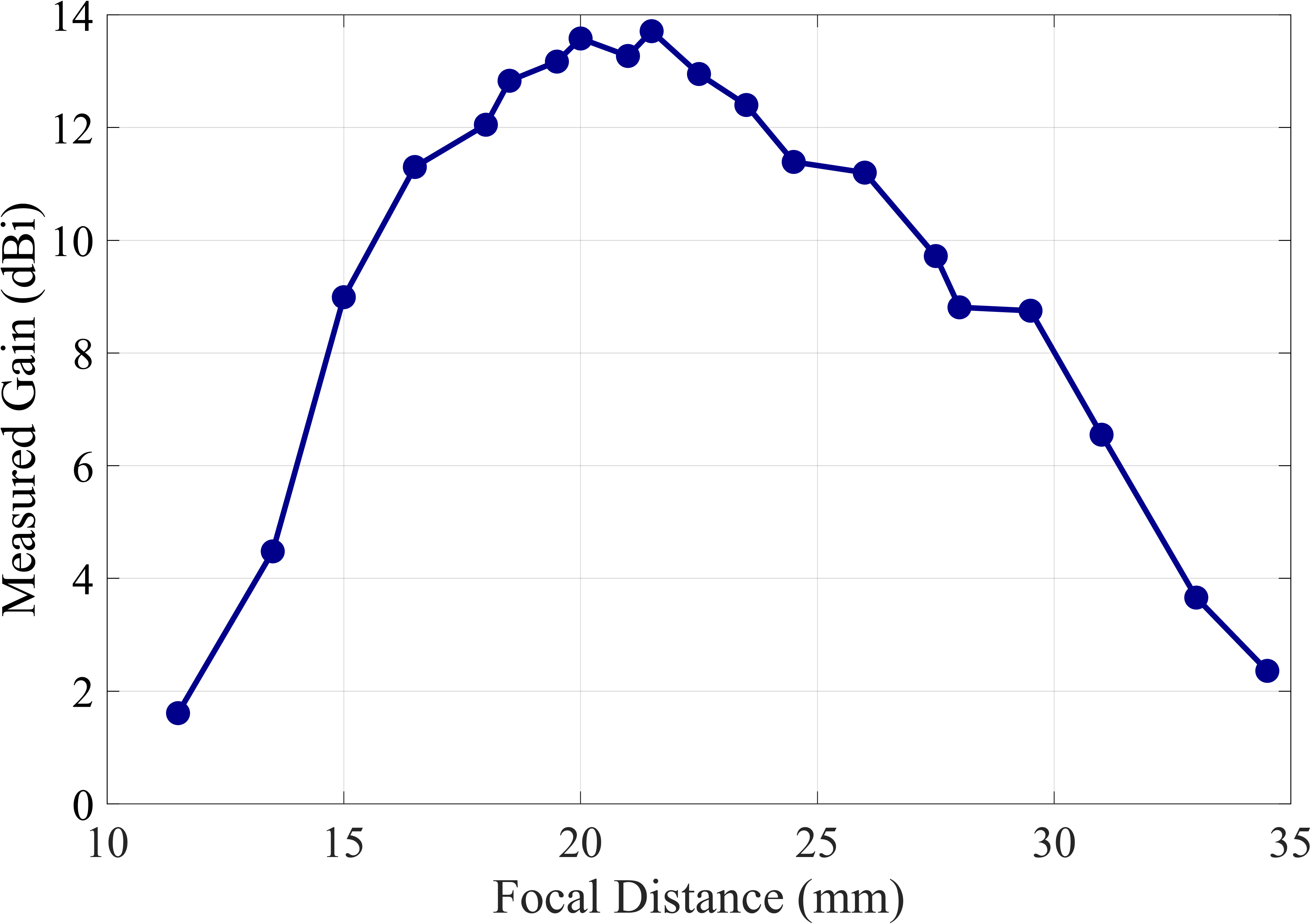}
        \caption{}
        \label{fig:Gain_Focalpoint_Result}
    \end{subfigure}
    \hspace*{0.01\columnwidth}
    \begin{subfigure}{0.32\columnwidth}
        \includegraphics[width=\textwidth]{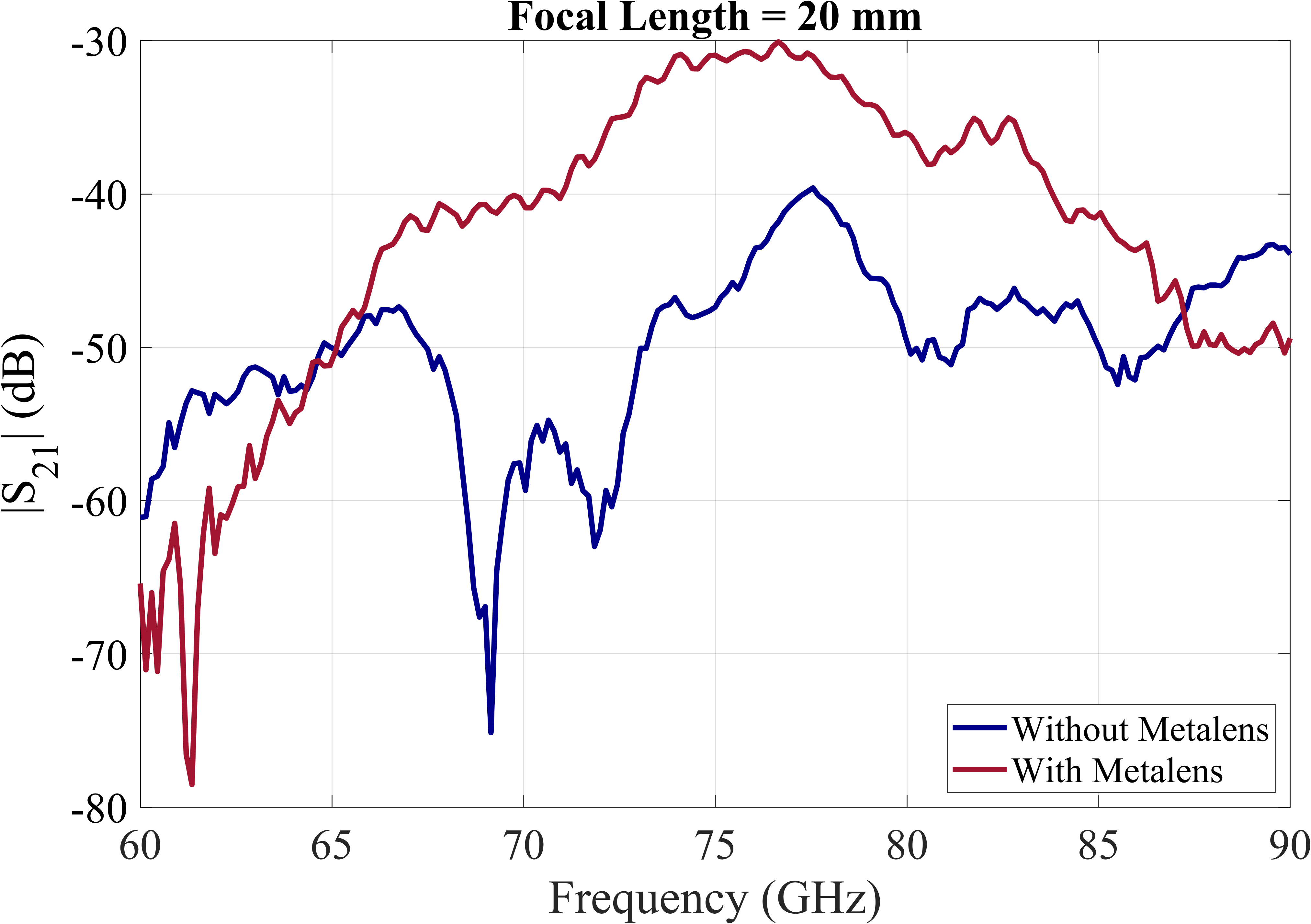}
        \caption{}
        \label{fig:S21_Frequency}
    \end{subfigure}

\caption{Metalens gain characterization: (a) measurement setup with the horn antenna (Port~1), the patch antenna (Port~2), and the metalens; (b) measured gain as a function of focal distance, showing peak gain at $20~\mathrm{mm}$; and (c) measured $|S_{21}|$ across the $60$--$90~\mathrm{GHz}$ band with and without the metalens at a fixed focal length of $20~\mathrm{mm}$.}
\label{fig:metalens_gain}
\vspace{-0.3cm}
\end{figure}

The metalens gain measurement setup is shown in Fig.~\ref{fig:Gain_Focalpoint_Setup}. A horn antenna connected to Port~1 of the VNA extender served as the transmitter, while a connectorized version of the patch antenna finalized in the ``Second Layer'' section was connected to Port~2 as the receiver. The metalens was mounted vertically on a plexiglass and foam support so that its center remained aligned with both antennas along the boresight axis. The focal distance, defined as the separation between the patch antenna and the metalens, was swept from approximately $10~\mathrm{mm}$ to $35~\mathrm{mm}$ in fine steps, and $S_{21}$ was recorded at each position. A reference measurement was then taken with the metalens removed. The realized gain of the metalens-based tag was extracted by subtracting the reference $S_{21}$ from the loaded $S_{21}$ and adding the assumed gain of the patch antenna ($5~\mathrm{dBi}$). The resulting gain of the metalens-based tag as a function of focal distance is presented in Fig.~\ref{fig:Gain_Focalpoint_Result}. The maximum peak gain of $13.58~\mathrm{dBi}$ occurs at a focal distance of approximately $20~\mathrm{mm}$, in close agreement with the focal length for which the metalens was designed.

To further characterize the metalens performance at the optimal focal distance, $S_{21}$ was measured across the $60$--$90~\mathrm{GHz}$ band with and without the metalens at a fixed focal length of $20~\mathrm{mm}$, as shown in Fig.~\ref{fig:S21_Frequency}. The metalens provides a clear gain enhancement throughout the operating band, with a peak improvement of $8.58~\mathrm{dB}$ near the design frequency of $78.5~\mathrm{GHz}$. This result is consistent with the gain enhancement of approximately $8.30~\mathrm{dB}$ observed at normal incidence in the RCS measurement section, confirming that the two independent measurement methods yield closely matching estimates of the metalens-induced gain enhancement.

\subsection*{RCS Measurements of Patch layer and Metalens-based tag}
\label{sec:RCS}
\begin{figure}[h]
\centering

% ---------- Row 1 ----------
\begin{subfigure}{0.25\linewidth}
\centering
\includegraphics[width=\linewidth]{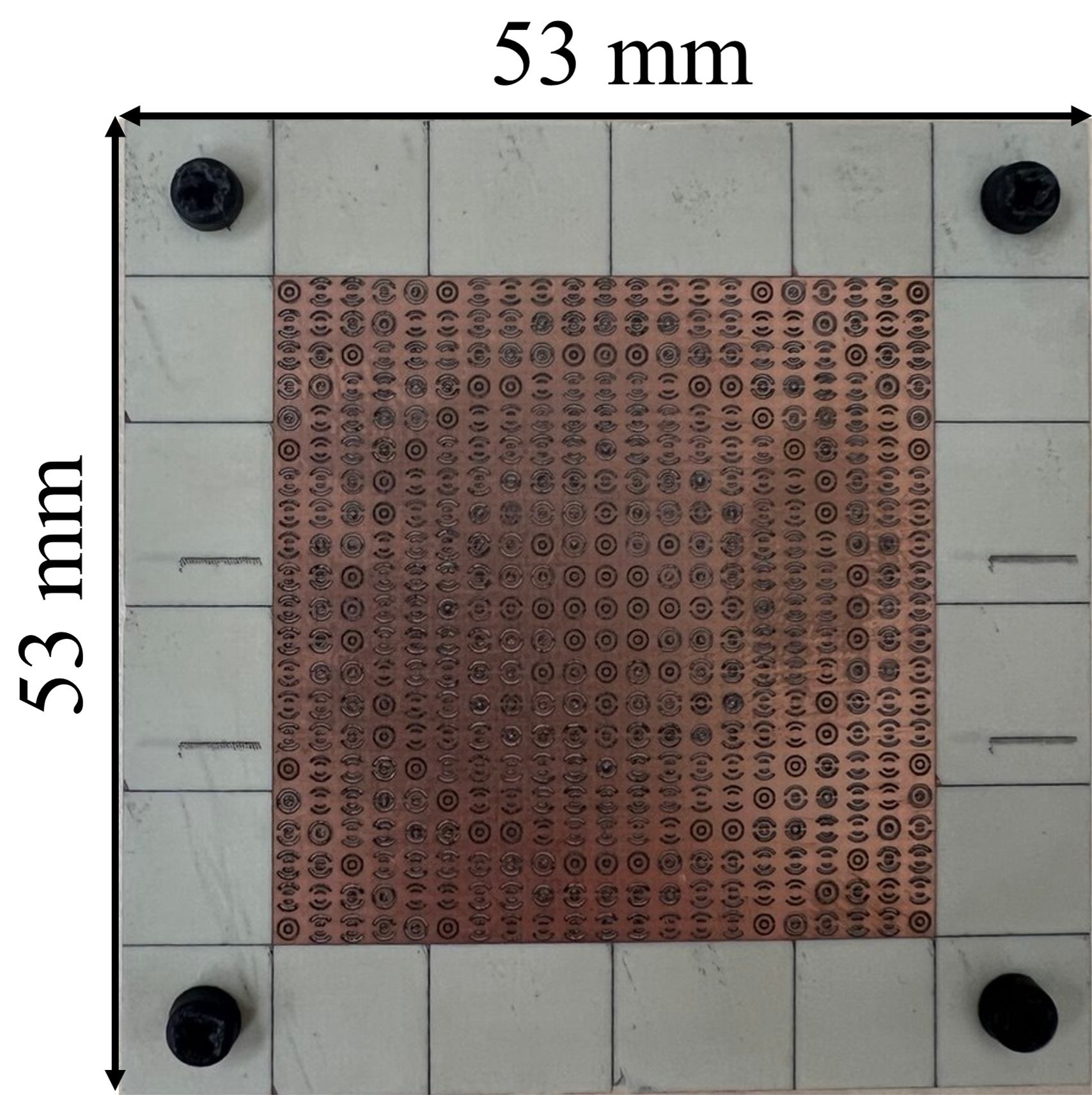}
\caption{}
\label{fig:front}
\end{subfigure}
\hspace*{0.03\columnwidth}
\begin{subfigure}{0.20\linewidth}
\centering
\includegraphics[width=\linewidth]{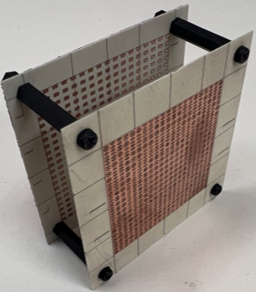}
\caption{}
\label{fig:MetalensPatch}
\end{subfigure}
\hspace*{0.03\columnwidth}
\begin{subfigure}{0.22\linewidth}
\centering
\includegraphics[width=\linewidth]{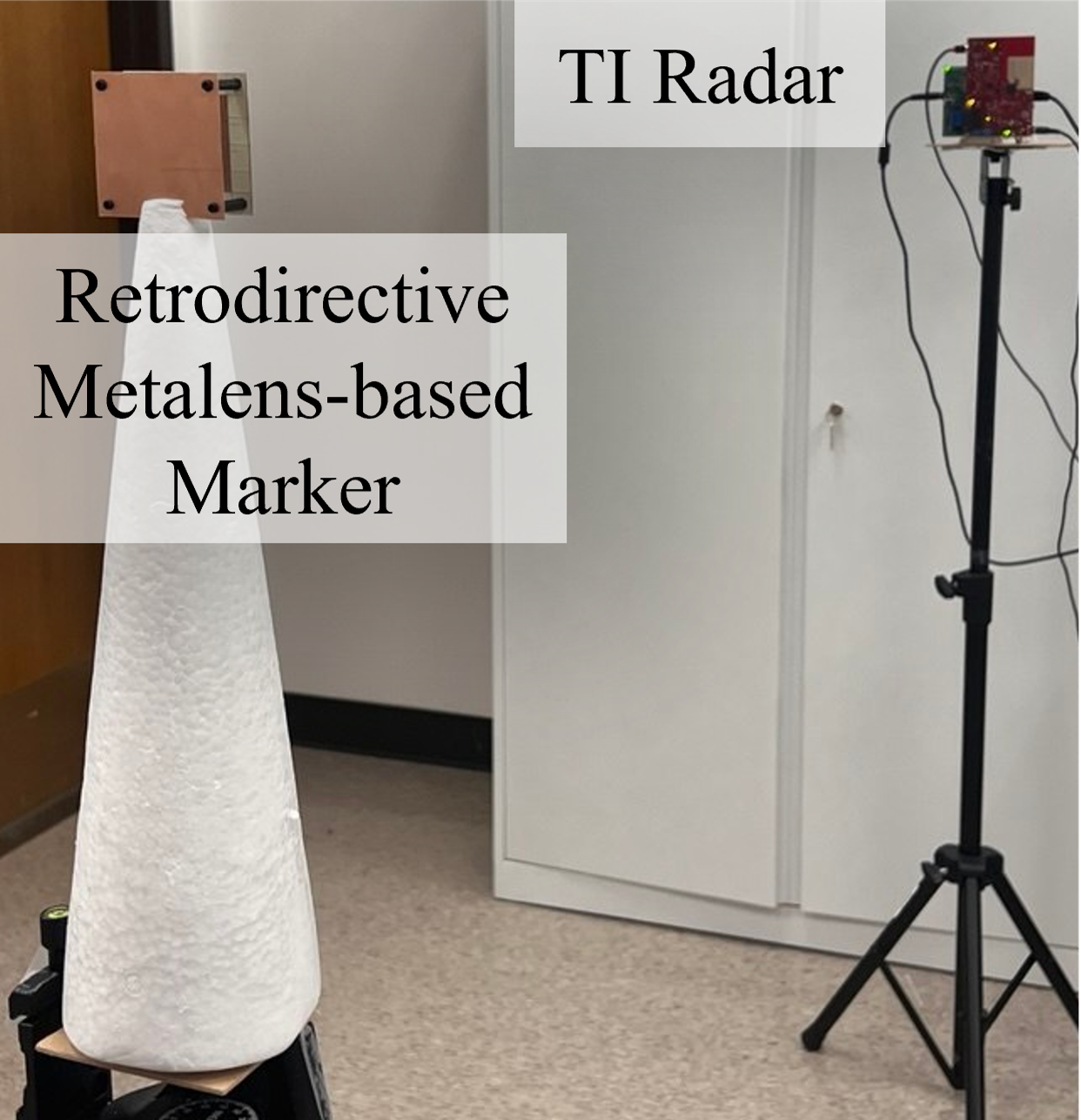}
\caption{}
\label{fig:RCS}
\end{subfigure}

\vspace{0.1cm}

% ---------- Row 2 ----------
\begin{subfigure}{0.4\linewidth}
\centering
\includegraphics[width=\linewidth]{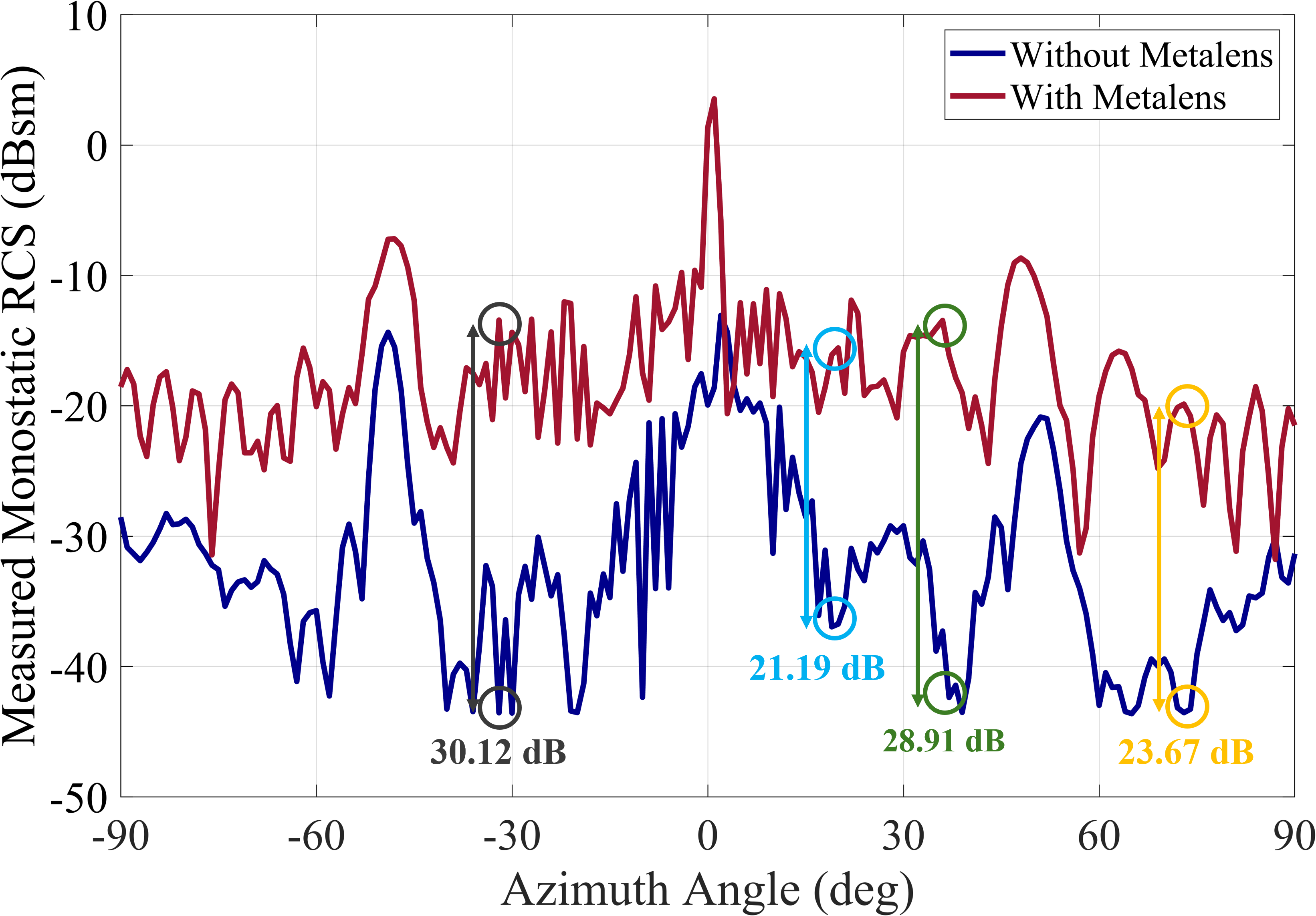}
\caption{}
\label{fig:RCS_Result}
\end{subfigure}
\hspace*{0.05\columnwidth}
\begin{subfigure}{0.4\linewidth}
\centering
\includegraphics[width=\linewidth]{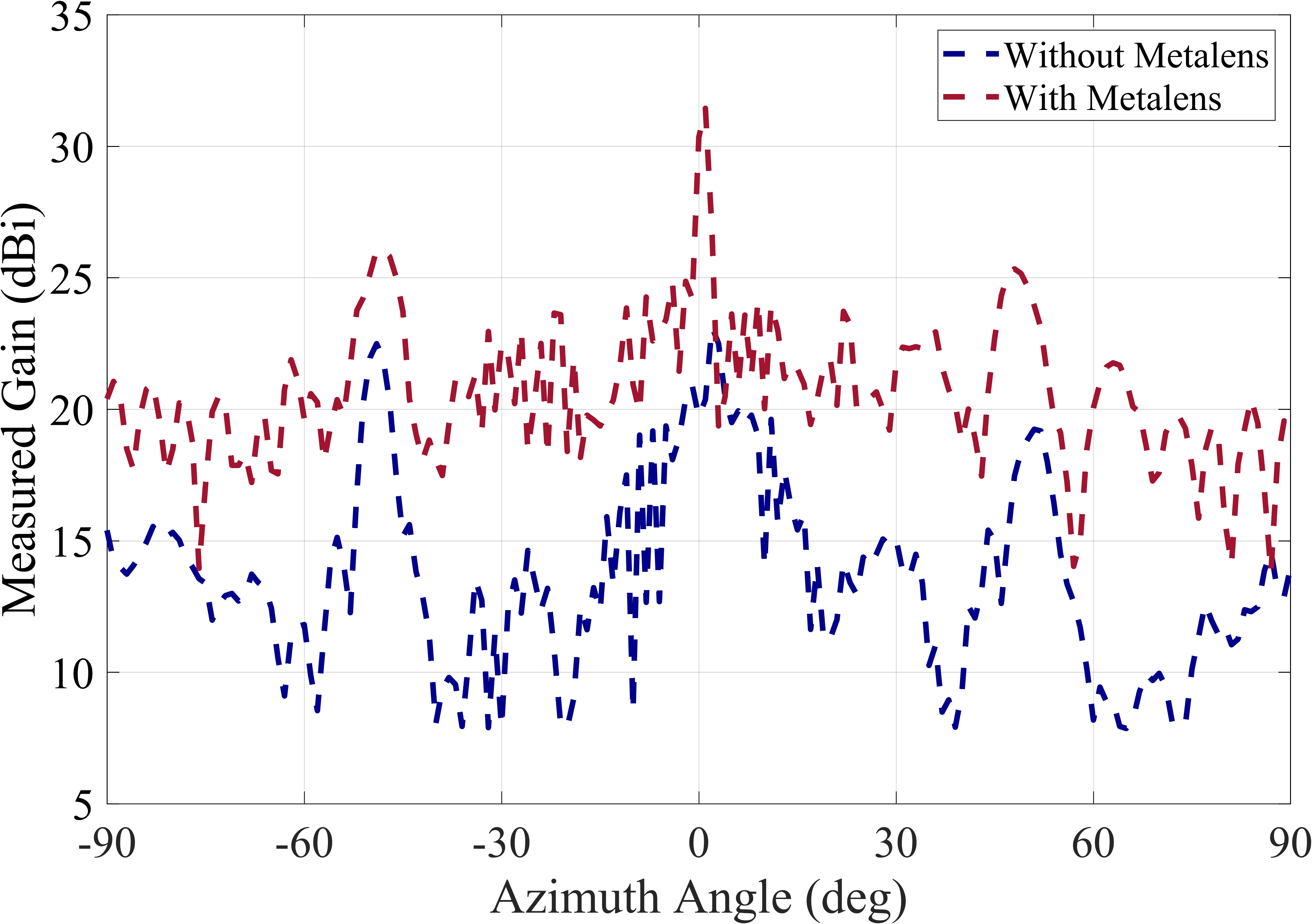}
\caption{}
\label{fig:Gain_Result}
\end{subfigure}

\caption{(a) Front view of the assembled tag. (b) Fabricated metalens and patch layer with $20~\mathrm{mm}$ spacing. (c) RCS measurement setup. (d) Comparison of measured RCS for the patch layer alone and for the complete metalens-based tag. (e) Corresponding gain comparison.}
\label{fig:RCS_related}
\vspace{-0.4cm}
\end{figure}

Both the patch antenna and metalens layers were fabricated using an LPKF ProtoLaser H4 PCB laser structuring system, which provides a structuring resolution of $75~\mathrm{\mu m}$. The two layers were assembled with an M2 standoff kit to maintain a $20~\mathrm{mm}$ separation, as shown in Fig.~\ref{fig:MetalensPatch}. To measure the RCS, we followed the same procedure described in \cite{JRFID} and performed the measurements using a Time-Division Multiplexing (TDM)-MIMO setup. The tag was mounted vertically on a non-reflective white foam support, as shown in Fig.~\ref{fig:RCS}, in order to generate vertical polarization consistent with that
of the radar. For the RCS measurements, the radar data were processed by computing the range FFT, from which the received power at the target location was extracted. The procedure was then repeated for $\phi$ angles ranging from $-90^\circ$ to $+90^\circ$ in $1^\circ$ increments. To calibrate and normalize the RCS measurements, a 2.175-inch diameter aluminum sphere was employed as a reference target. Its theoretical RCS is given by $\sigma = \pi a^2$, where $a$ is the sphere radius, yielding approximately $-26.19~\mathrm{dBsm}$ for the chosen size. The sphere was mounted on the same white foam structure used for the tag to ensure consistent conditions. The radar return from the sphere was then compared with its known RCS to determine a calibration factor using the radar range equation. This factor remains constant as long as the measurement range is unchanged. Once established, it allows the received power from any target placed at the same distance to be directly mapped to its RCS. Following the same procedure described above, the RCS was first measured for the patch layer alone, without the metalens, and then for the complete retrodirective metalens-based tag under identical conditions. The results for both cases are shown in Fig.~\ref{fig:RCS_Result}, with peak RCS values of $3.54~\mathrm{dBsm}$ for the metalens-based tag and $-13.06~\mathrm{dBsm}$ for the patch layer only at normal incidence, corresponding to an RCS enhancement of $16.60~\mathrm{dB}$. Using these RCS values, the corresponding equivalent gain was computed based on the monostatic RCS--gain relation, $G = (\sigma_{\mathrm{dBsm}} - 20\log_{10}(\lambda) + 10\log_{10}(4\pi))/2$, where $\sigma_{\mathrm{dBsm}}$ is the measured monostatic RCS and $\lambda$ is the wavelength. The resulting gain is shown in Fig.~\ref{fig:Gain_Result}, with peak values of $31.44~\mathrm{dBi}$ for the metalens-based tag and $23.14~\mathrm{dBi}$ for the patch layer only at normal incidence, corresponding to a gain improvement of $8.30~\mathrm{dB}$.

Since the patch layer is uniform and implemented on a metallic ground plane, it shows strong reflection near broadside due to a combination of structural-mode scattering and antenna-mode re-radiation. However, as the incident angle increases, these reflections are directed towards the specular direction, resulting in a rapid degradation of the power backscattered towards the source, as shown in Fig.~\ref{fig:RCS_related}. As the illumination angle further increases to approximately $\pm 50^\circ$, grating lobes arising from the periodicity of the patch antenna layer appear to provide high detectability once again. The patch layer has a center-to-center period of $d = 2.48~\mathrm{mm}$. Under monostatic illumination, the round-trip path difference between adjacent patches is $2d\sin\theta$, so constructive backscatter from the periodic patch layer occurs when $2d\sin\theta = m\lambda$. At $78.5~\mathrm{GHz}$ ($\lambda = 3.82~\mathrm{mm}$), the first-order ($m = 1$) Bragg lobes are predicted at $\theta = \pm\arcsin(\lambda/(2d)) = \pm 50.4^\circ$, matching the peaks observed in the measured RCS. In contrast, the metalens-based tag maintains strong retroreflection over a $\pm 40^\circ$ angular range. At $20^\circ$, the RCS improves by approximately $21.19~\mathrm{dB}$ compared to the patch-only case, and this enhancement becomes more pronounced at larger angles. On average, the metalens-based tag achieves a gain improvement of $7.58~\mathrm{dB}$ and an RCS enhancement of $15.16~\mathrm{dB}$ over the entire angular space from $-40^\circ$ to $40^\circ$. At larger deviations from normal incidence ($19^\circ$ to $40^\circ$ and $-40^\circ$ to $-19^\circ$), the improvements reach $9.42~\mathrm{dB}$ in gain and $18.84~\mathrm{dB}$ in RCS. Therefore, the metalens stabilizes the RCS response, exhibiting approximately $10~\mathrm{dB}$ variation with respect to the average RCS over an $80^\circ$ angular range, whereas the patch-only structure suffers from a much larger RCS variation, up to $30~\mathrm{dB}$ drop away from broadside. These results demonstrate that the metalens effectively suppresses structural-mode nulls and enables wide-angle retrodirective performance. A comparison with prior passive radar markers that report measured RCS in the automotive frequency band is summarized in Table~\ref{tab:comparison}. The reflector in~\cite{douglas2023low} achieves a median RCS of $-19~\mathrm{dBsm}$ over its $5^{\circ}$ of azimuthal coverage. The design in~\cite{IEEERFID2024} provides $80^{\circ}$ of angular coverage; however, its median RCS of $-30~\mathrm{dBsm}$ limits the detection range to only $11~\mathrm{m}$, and its horizontal polarization is incompatible with the vertically polarized radars used in commercial vehicles. The SIW horn Van~Atta design in~\cite{JRFID} achieves a higher median RCS of $-21~\mathrm{dBsm}$, but its Van~Atta-based architecture faces limitations for further scaling the RCS. Building on these findings, the retrodirective radar marker proposed in this work achieves a median RCS of $-16.76~\mathrm{dBsm}$ across $80^{\circ}$ of angular coverage, simultaneously delivering the highest detectability and the widest field of view among the compared markers, while maintaining a compact form factor suitable for cyclist applications and thereby contributing to improved safety for vulnerable road users. Moreover, compared to the patch layer alone, the addition of the metalens improves the median RCS by $9.92~\mathrm{dB}$, demonstrating its effectiveness.

\renewcommand{\arraystretch}{0.8}
\begin{table}[h]
\centering
\begin{tabular}{|c|c|c|c|}
\hline
Ref. & Size (mm$^3$) & Median RCS & Angular Coverage \\
\hline
\cite{douglas2023low} 
& $64.6 \times 44.2 \times 0.2$ 
& $-19$ dBsm 
& $5^{\circ}$ \\
\hline
\cite{IEEERFID2024} 
& $18.54 \times 12.95 \times 12.4$ 
& $-30$ dBsm 
& $80^{\circ}$ \\
\hline
\cite{JRFID} 
& $56 \times 40.76 \times 6.54$ 
& $-21$ dBsm 
& $30^{\circ}$ \\
\hline
Patch layer 
& $53 \times 53 \times 0.035$ 
& $-26.68$ dBsm 
& $40^{\circ}$ \\
\hline
This work 
& $53 \times 53 \times 20$ 
& $-16.76$ dBsm 
& $80^{\circ}$ \\
\hline
\end{tabular}
\caption{\label{tab:comparison}
Performance comparison of the proposed tag with previously reported radar markers.}
%\vspace{-0.2cm}
\end{table}

\subsection*{Detection Range of the Metalens-based Tag }

Measurements were performed using the AWR2944EVM automotive radar from Texas Instruments~\cite{ti_awr2944evm}. The radar was configured to chirp from $76.81~\mathrm{GHz}$ to $81~\mathrm{GHz}$, resulting in a bandwidth of $4.19~\mathrm{GHz}$ and a corresponding range resolution of $3.58~\mathrm{cm}$. Each chirp lasted $430~\mu\mathrm{s}$, with data sampling starting $6~\mu\mathrm{s}$ after the chirp onset. The slope was $10.235~\mathrm{MHz}/\mu\mathrm{s}$, resulting in a maximum detectable range of approximately $75~\mathrm{m}$. A total of 4096 samples were acquired per chirp at a sampling rate of $10~\mathrm{Msps}$. The radar features four transmit (Tx) and four receive (Rx) channels, with MIMO functionality enabled using TDM. In TDM MIMO, separate time slots were assigned to each transmit antenna, ensuring orthogonal signaling and preventing interference. This allowed the receiver antennas to distinguish each transmitter’s signal based on its corresponding time slot. Each frame was divided into multiple blocks, with each block consisting of four time slots, one for each transmit antenna (Tx). Accordingly, the chirp sequence length matched the number of transmitters (four in this case) and was repeated 128 times per frame. To extract the range and angle of arrival (AoA) of the tag response, the received data from each chirp was processed using a two-stage Fast Fourier Transform (FFT). First, a range FFT was applied along the fast-time dimension of each chirp. Then, an angular FFT was performed across the received antenna channels to resolve the AoA. This procedure was repeated for all chirps in the frame, and the resulting 2D range–angle spectra were magnitude-averaged across chirps. The tag response was identified as the strongest peak in the averaged 2D spectrum, and its amplitude was recorded.

\begin{figure}[!t]
    \centering
 \begin{subfigure}{0.2\columnwidth}
       \includegraphics[{width=\textwidth}]{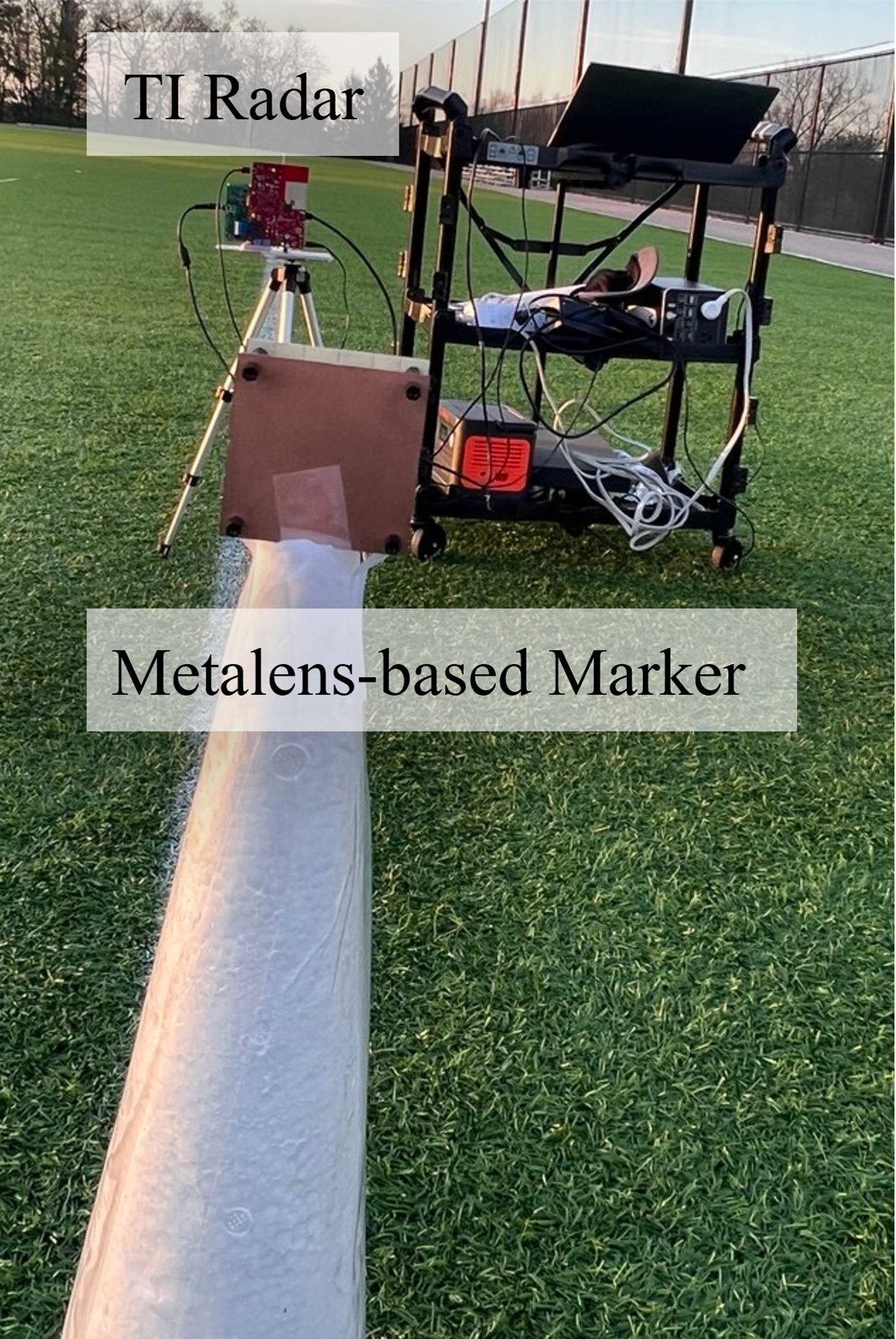}
       \caption{}
       \label{fig:SNR_setup_outdoor}
     \end{subfigure}
 \hspace*{0.01\columnwidth}
       \begin{subfigure}{0.45\columnwidth}
       \includegraphics[width=\textwidth]{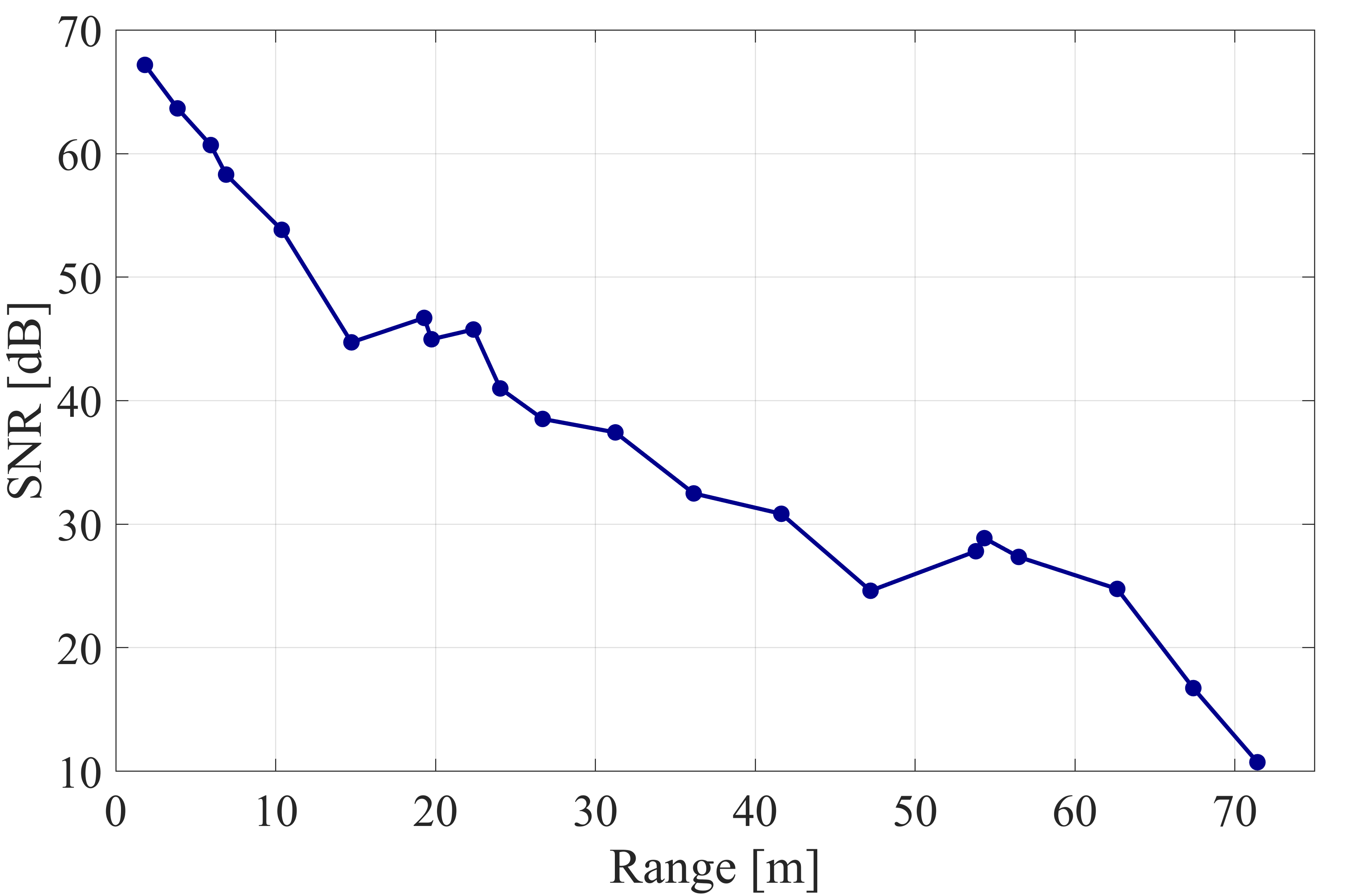}
       \caption{}
       \label{fig:SNR_Result}
       \end{subfigure}

     \caption{Outdoor SNR measurement of the proposed metalens-based tag: (a) experimental setup, and (b) measured SNR as a function of distance.}
	\label{fig:SNR}
 \vspace{-0.2cm}
	\end{figure}

To evaluate the range detection capability of the proposed tag, the signal-to-noise ratio (SNR) was measured at multiple distances. The signal level was obtained from the power spectrum at the range and angular bins corresponding to the tag response, while the noise level was estimated by averaging the power spectrum when the radar was pointed toward the sky. The SNR was then obtained by subtracting the noise level from the signal level in decibels. Fig.~\ref{fig:SNR_setup_outdoor} shows the experimental setup for the SNR measurements in an outdoor environment. The radar was placed on a tripod, and the tag was mounted on non-reflective white foam, tilted by approximately $5^\circ$ in azimuth to minimize ground reflections. The radar and tag were placed at the same height. Fig.~\ref{fig:SNR_Result} shows the measured SNR of the proposed metalens-based tag versus distance. The radar detected the tag at a range of $71.41~\mathrm{m}$ with an SNR of $10.73~\mathrm{dB}$, exceeding the 10~dB threshold commonly used for reliable detection. This result demonstrates the suitability of the proposed structure for long-range outdoor sensing and its potential for practical real-world applications.

\subsection*{Imaging Performance of the Metalens-Based Marker}

Metallic objects with curved or cylindrical surfaces, such as bicycle frames, scatter incident energy in many directions rather than reflecting it back toward the radar, producing weak monostatic backscatter. Their radar return is also subject to scintillation, further compromising the reliability of radar-based detection in realistic scenarios. Retrodirective markers, in contrast, rely primarily on their antenna-mode response rather than structural scattering, and are capable of reflecting the incident signal back toward the radar even at oblique angles. The goal of this imaging experiment is to evaluate the ability of the proposed metalens-based marker to enhance the radar image of a metallic object at both normal and oblique azimuthal angles.

\vspace{0.1cm}

\begin{figure*}[h]
    \centering
    \begin{subfigure}{0.18\textwidth}
        \centering
        \includegraphics[width=\linewidth]{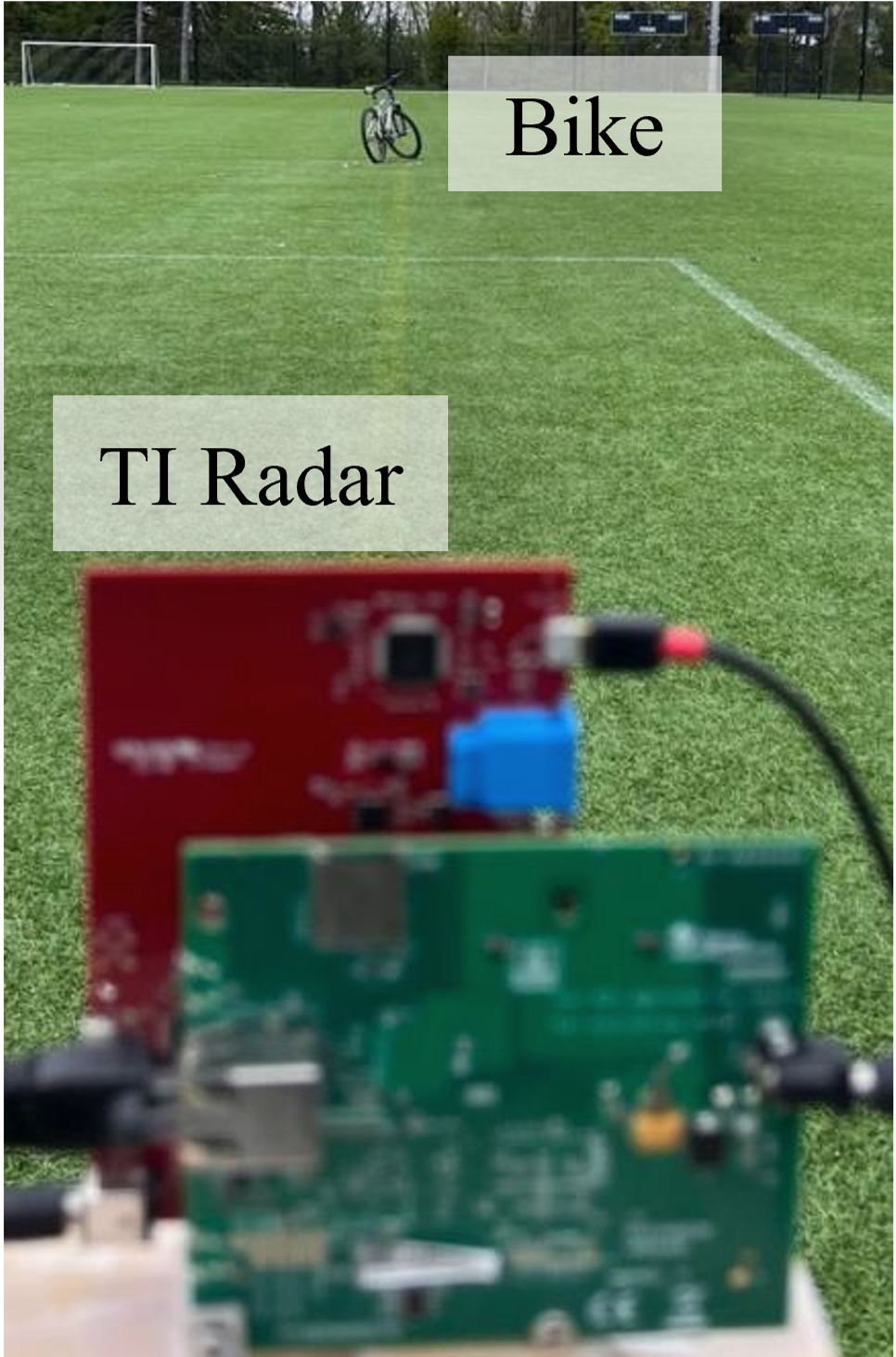}
        \caption{}
        \label{fig:setup_a}
    \end{subfigure}
    \hspace{2em}
    \begin{subfigure}{0.31\textwidth}
        \centering
        \includegraphics[width=\linewidth]{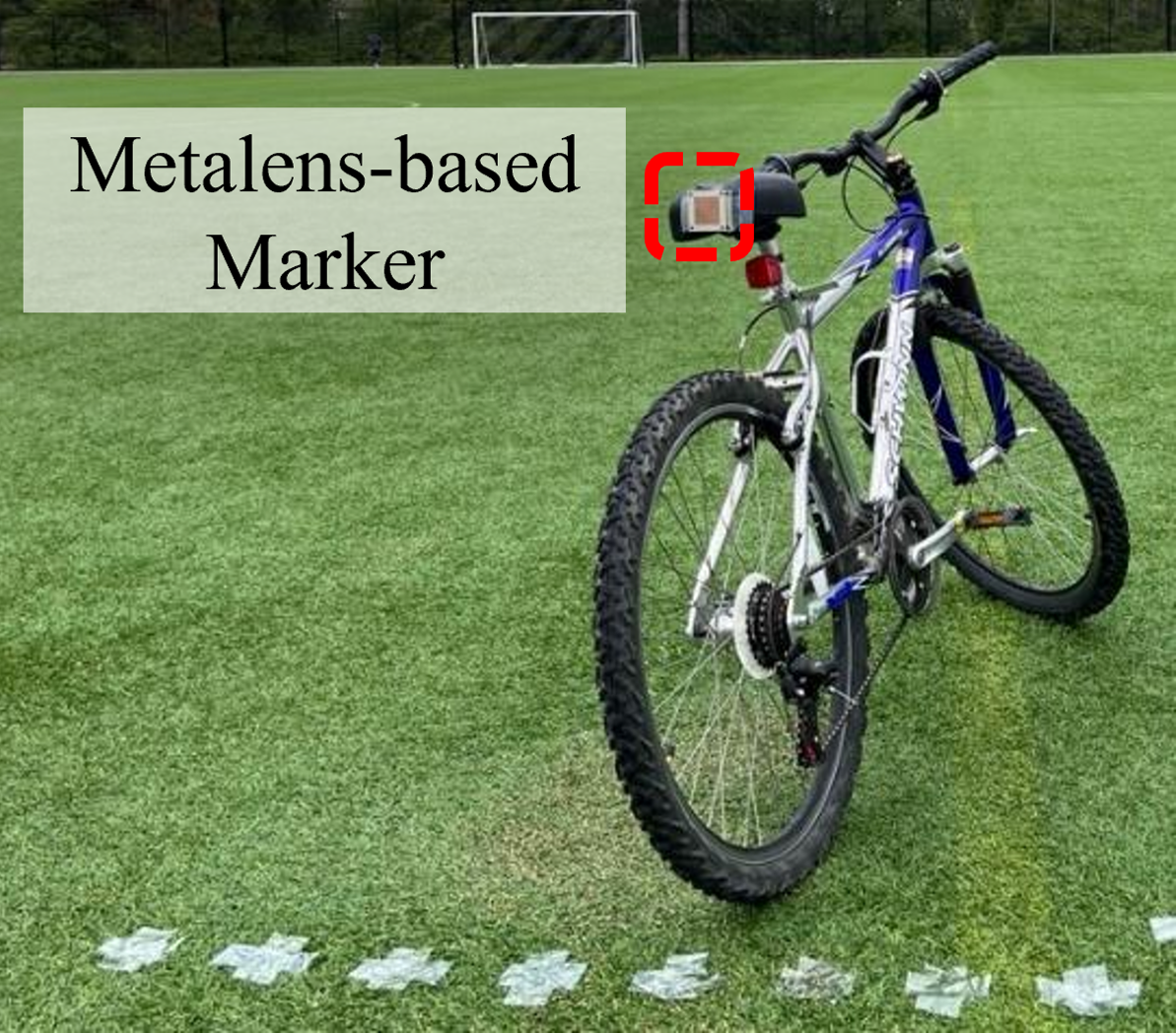}
        \caption{}
        \label{fig:setup_b}
    \end{subfigure}
    \hspace{2em}
    \begin{subfigure}{0.221\textwidth}
        \centering
        \includegraphics[width=\linewidth]{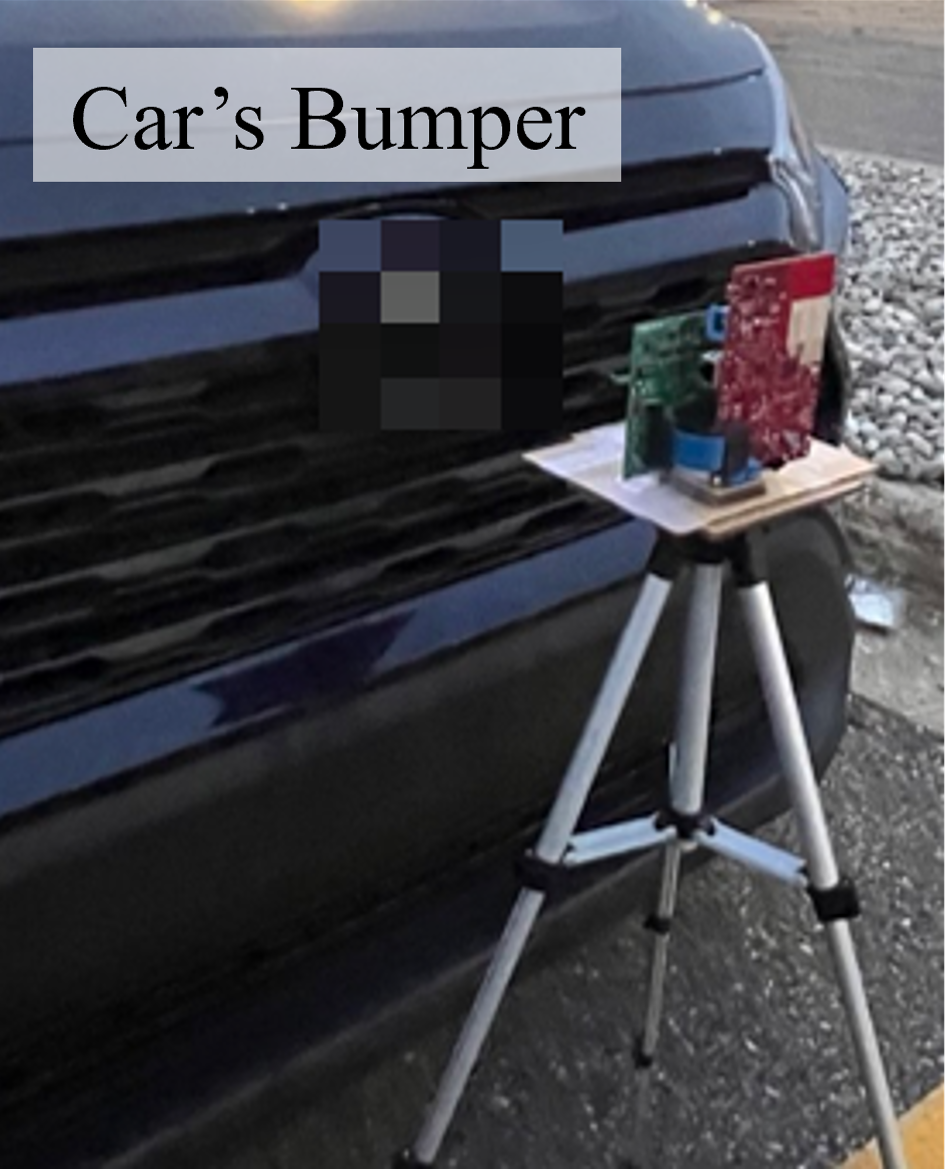}
        \caption{}
        \label{fig:setup_c}
    \end{subfigure}
    \caption{Outdoor imaging setup for evaluating the metalens-based marker's performance on a bicycle. (a) The bicycle is placed in front of the TI mmWave radar approximately $20~\mathrm{m}$ away. (b) Close-up of the metalens-based marker mounted on the back seat of the bicycle. (c) Radar mounted on a tripod at the same height as a typical automotive radar installation, placed in front of a car to emulate a vehicle-mounted sensing configuration.}
    \label{fig:imaging_setup}
 \vspace{-0.1cm}    
\end{figure*}

\begin{figure*}[h]
    \centering
    \begin{subfigure}{0.24\textwidth}
        \includegraphics[width=\linewidth]{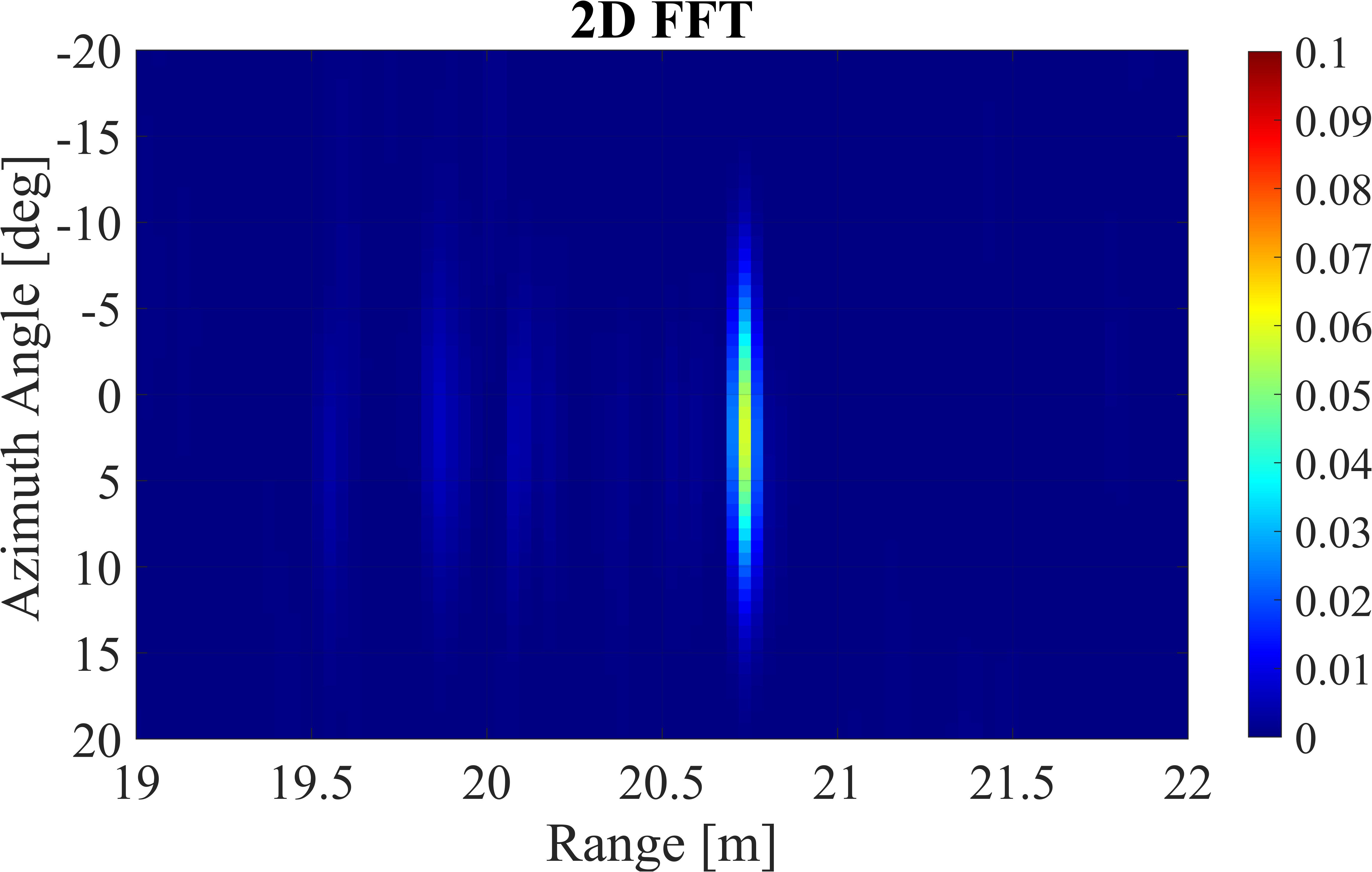}
        \caption{$0^\circ$, no marker}
        \label{fig:a}
    \end{subfigure}
    \hfill
    \begin{subfigure}{0.24\textwidth}
        \includegraphics[width=\linewidth]{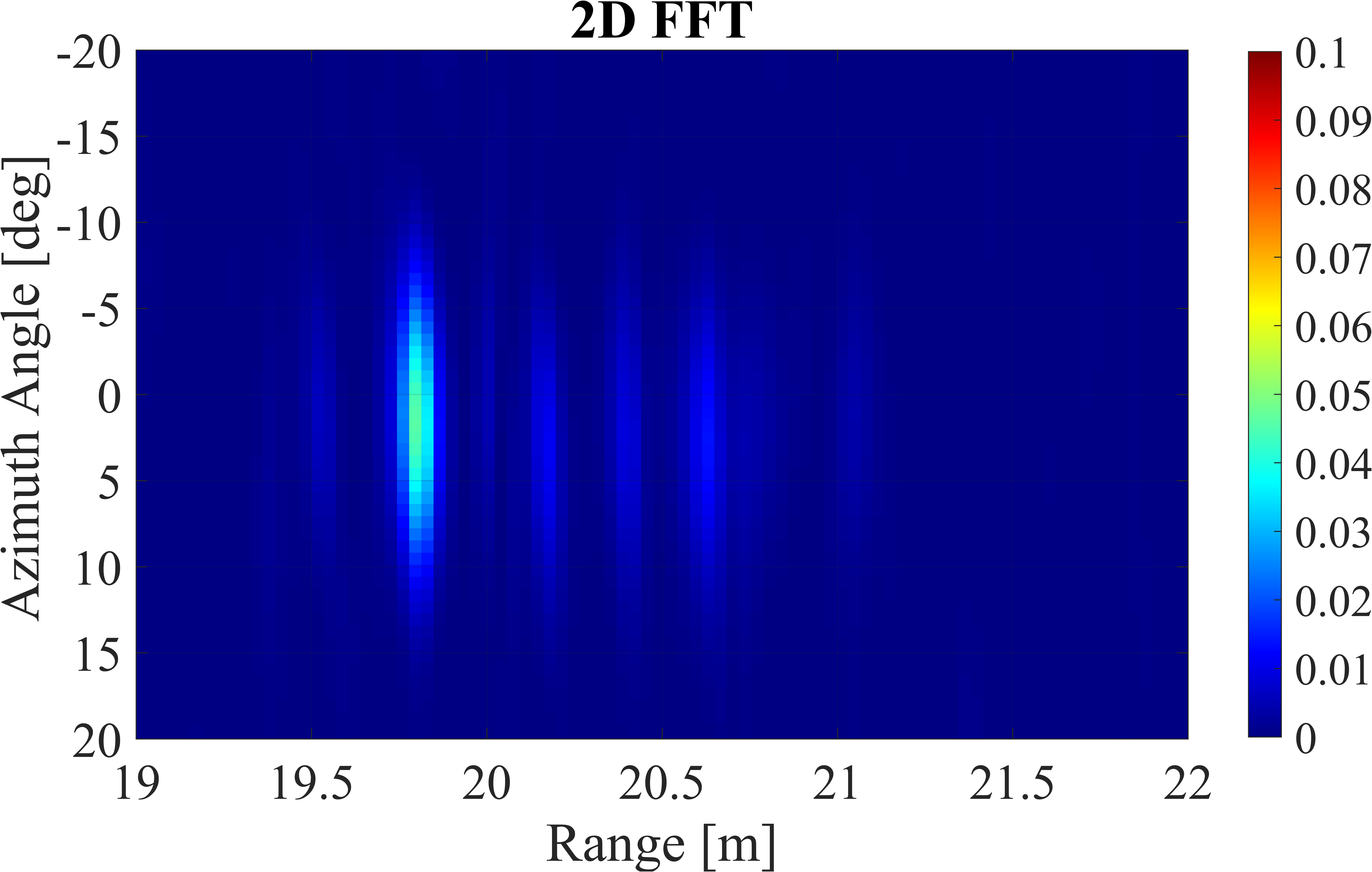}
        \caption{$5^\circ$, no marker}
        \label{fig:b}
    \end{subfigure}
    \hfill
    \begin{subfigure}{0.24\textwidth}
        \includegraphics[width=\linewidth]{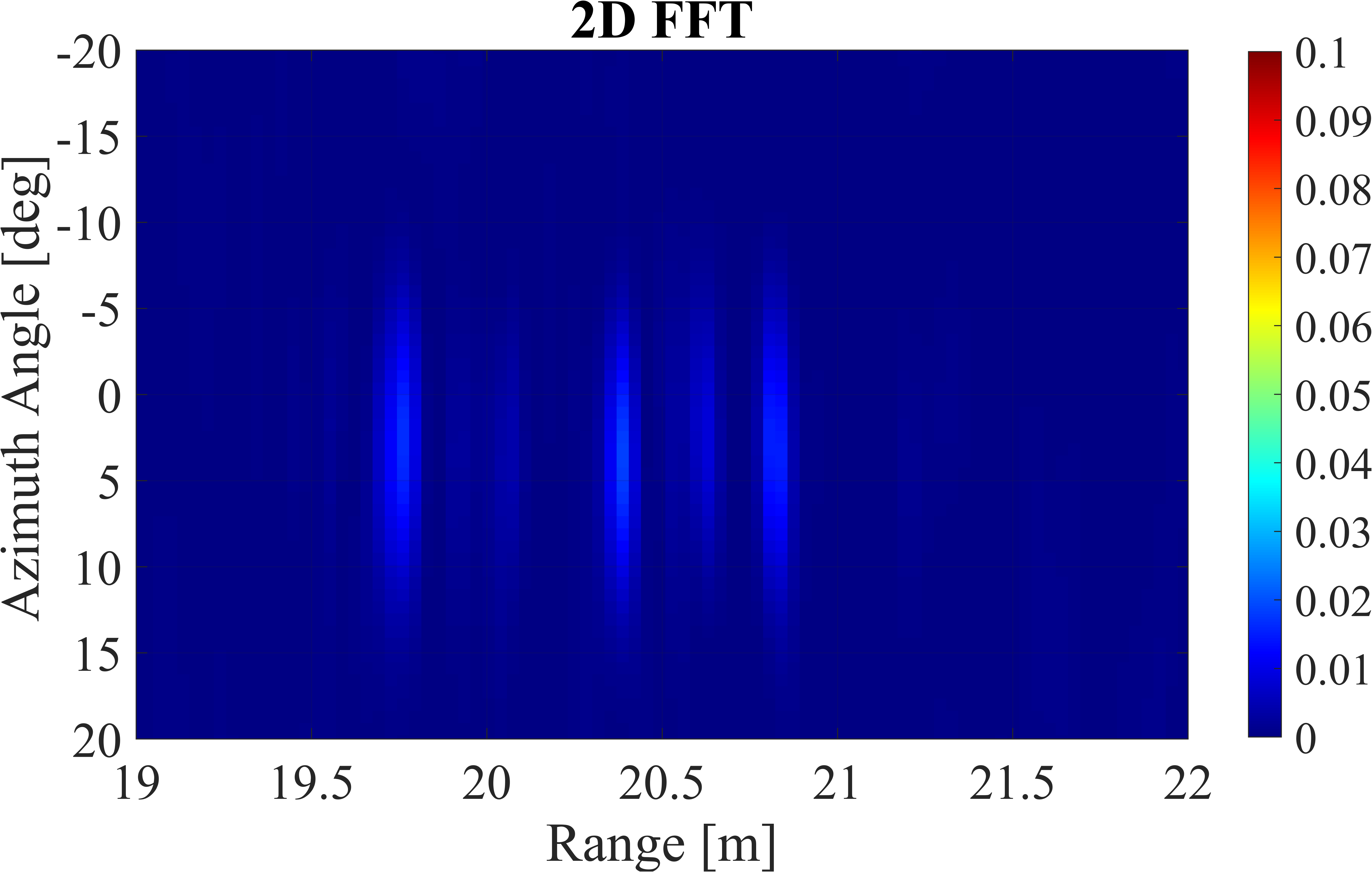}
        \caption{$10^\circ$, no marker}
        \label{fig:c}
    \end{subfigure}
    \hfill
    \begin{subfigure}{0.24\textwidth}
        \includegraphics[width=\linewidth]{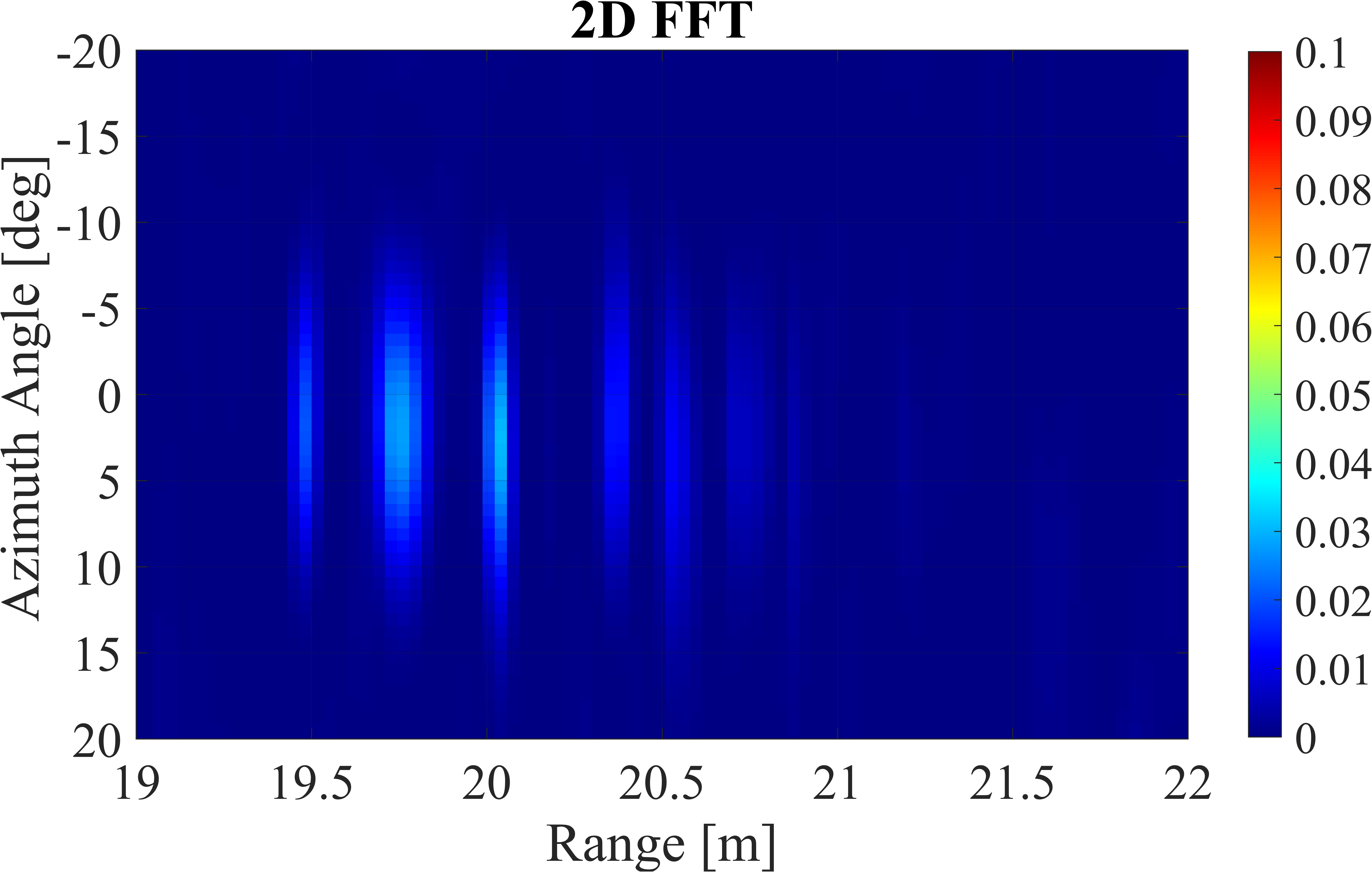}
        \caption{$15^\circ$, no marker}
        \label{fig:d}
    \end{subfigure}
    
    \vspace{0.5em}
    
    \begin{subfigure}{0.24\textwidth}
        \includegraphics[width=\linewidth]{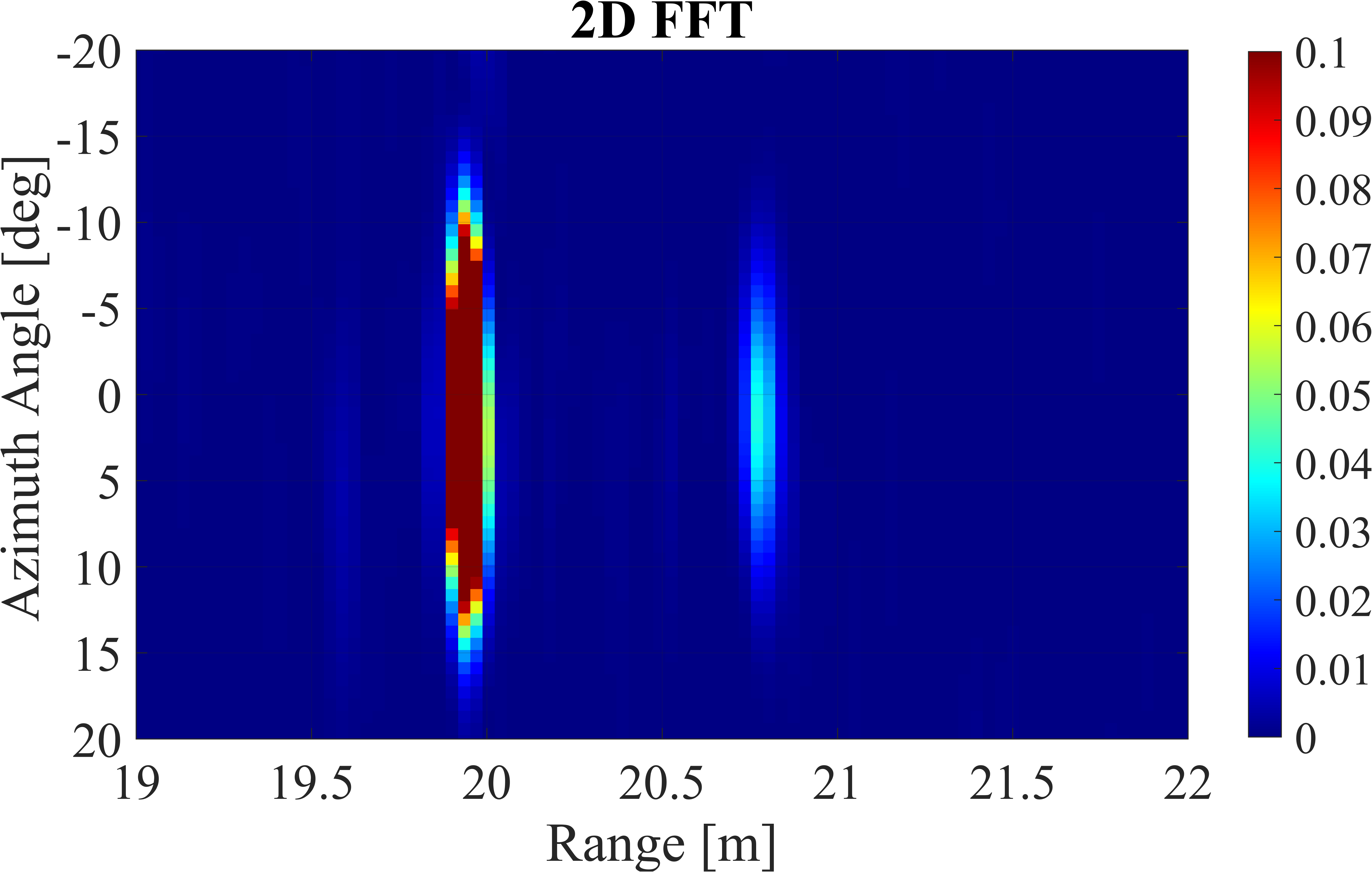}
        \caption{$0^\circ$, with marker}
        \label{fig:e}
    \end{subfigure}
    \hfill
    \begin{subfigure}{0.24\textwidth}
        \includegraphics[width=\linewidth]{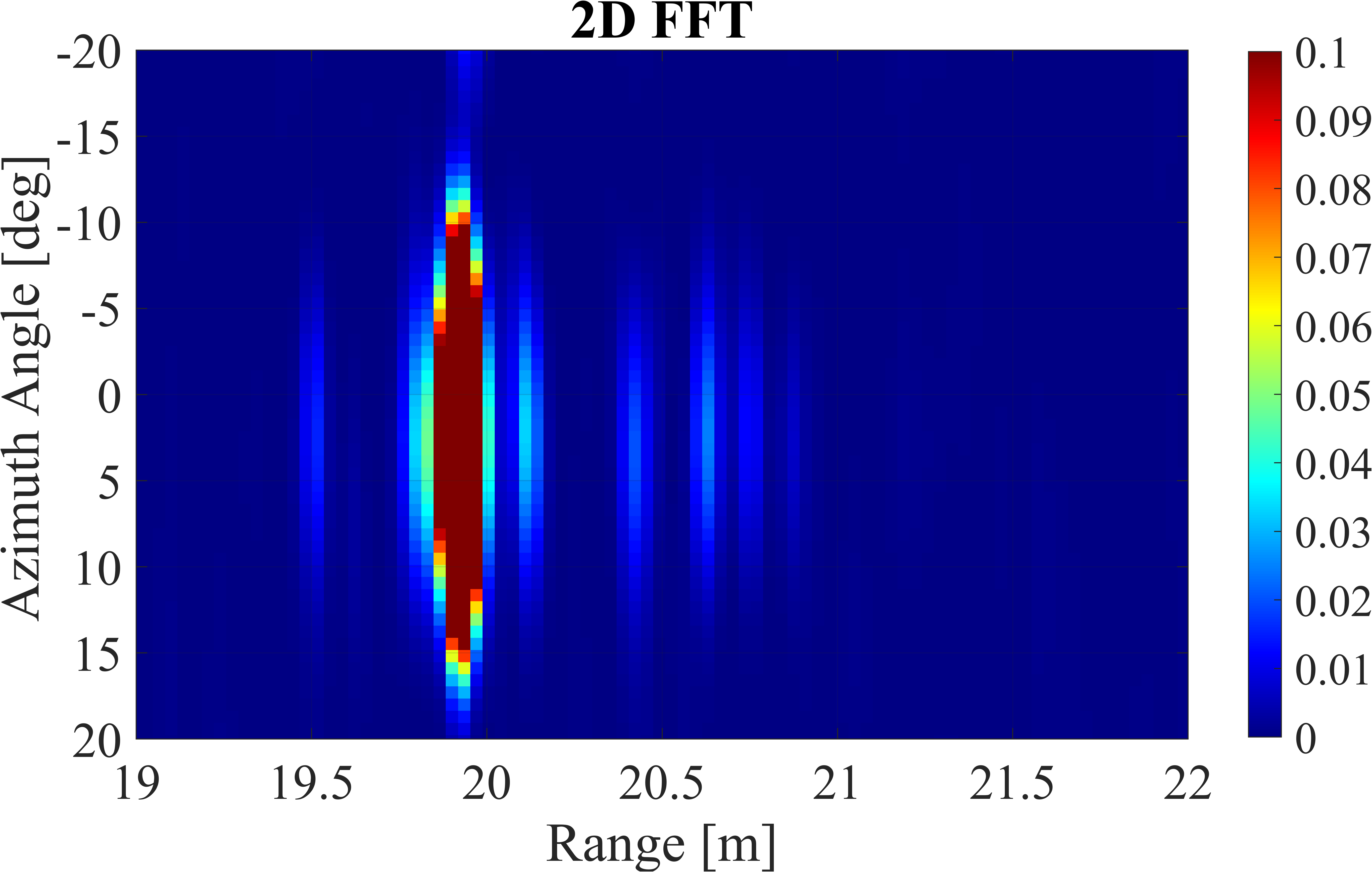}
        \caption{$5^\circ$, with marker}
        \label{fig:f}
    \end{subfigure}
    \hfill
    \begin{subfigure}{0.24\textwidth}
        \includegraphics[width=\linewidth]{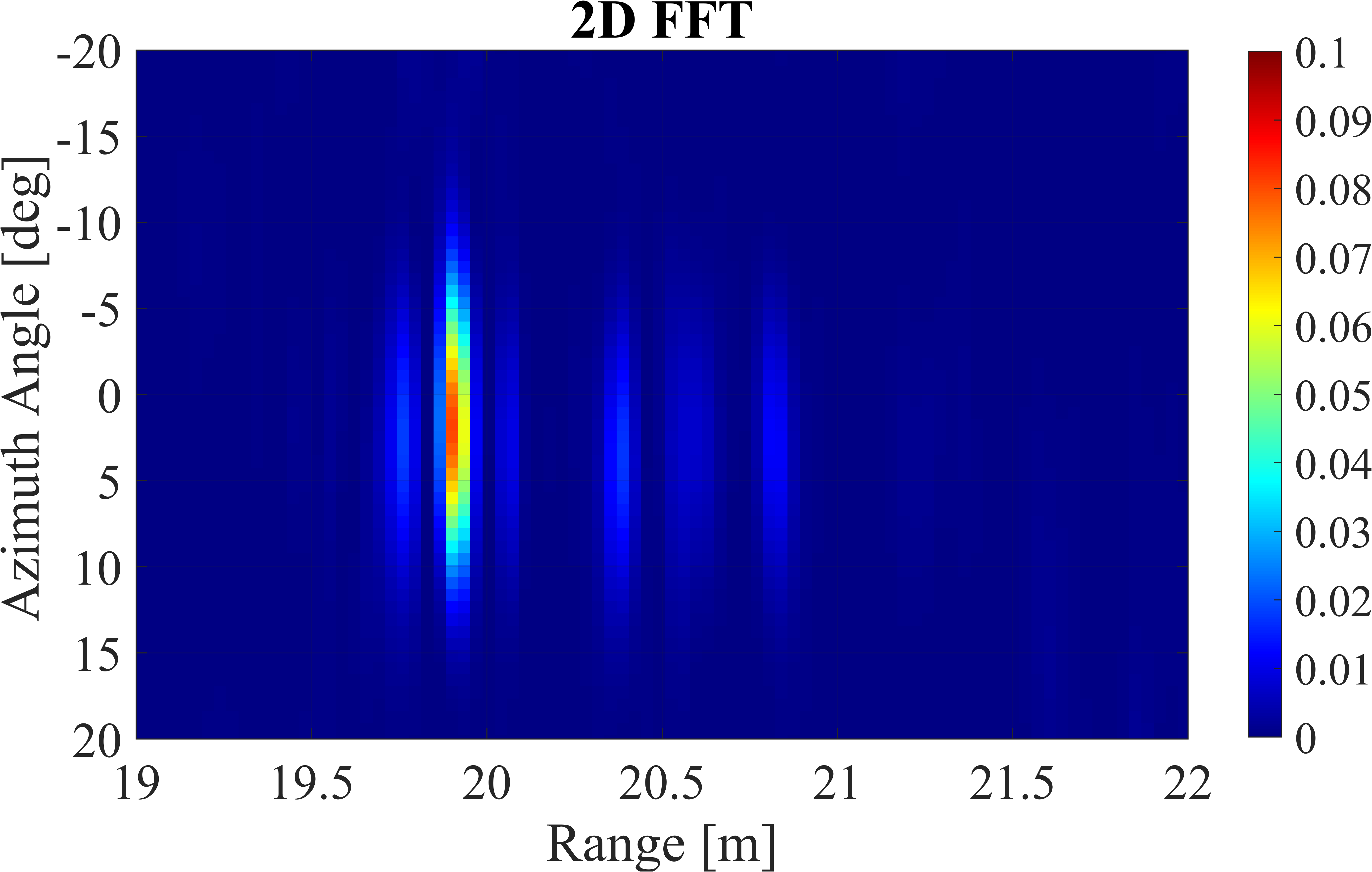}
        \caption{$10^\circ$, with marker}
        \label{fig:g}
    \end{subfigure}
    \hfill
    \begin{subfigure}{0.24\textwidth}
        \includegraphics[width=\linewidth]{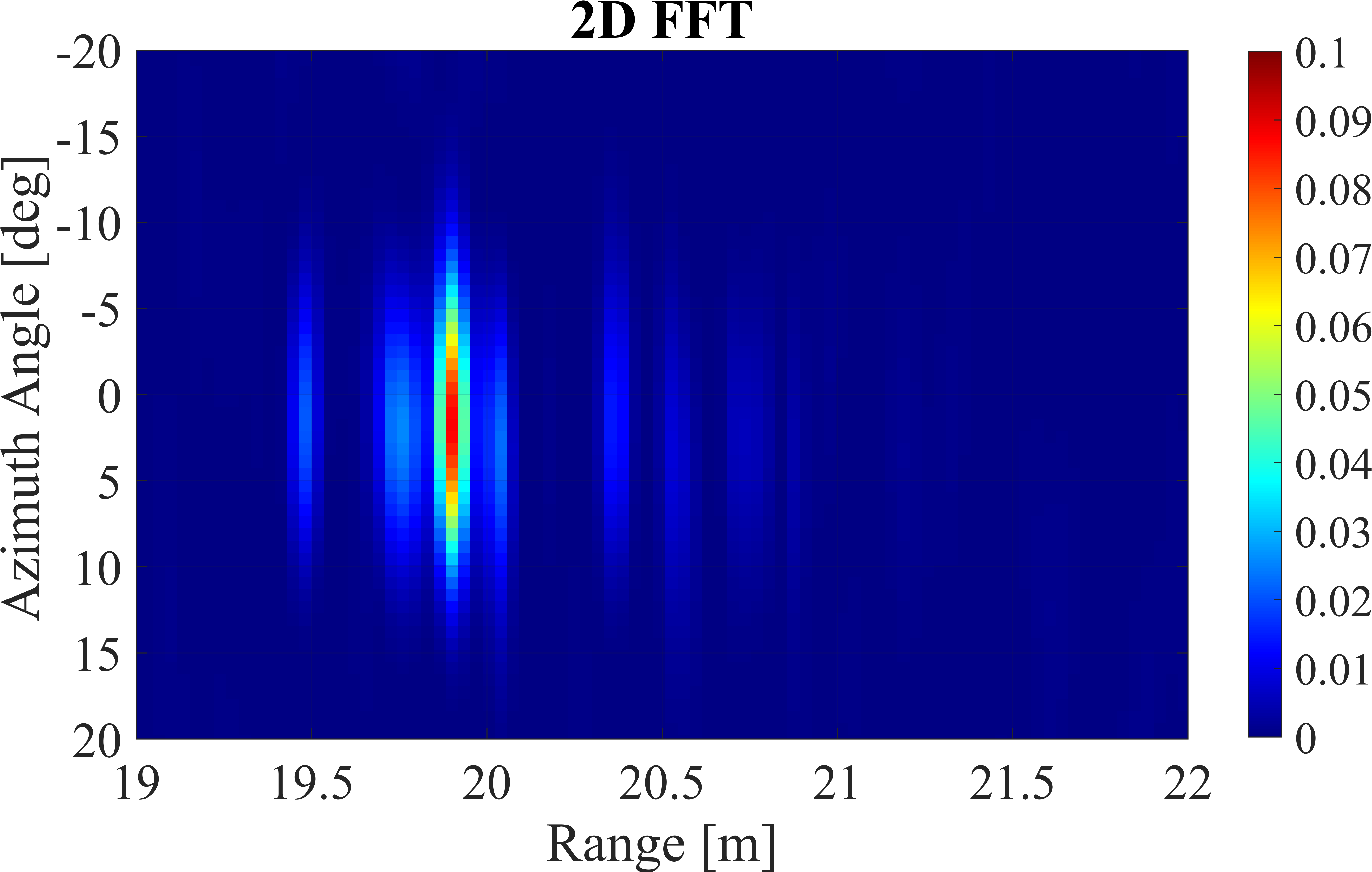}
        \caption{$15^\circ$, with marker}
        \label{fig:h}
    \end{subfigure}
    
    \caption{Range--azimuth maps obtained from the 2D FFT for the bicycle at four azimuthal angles ($0^\circ$, $5^\circ$, $10^\circ$, $15^\circ$). \textbf{Top row:} bicycle without the metalens-based marker. \textbf{Bottom row:} bicycle with the marker. The bare bicycle produces a weak return that further degrades at oblique angles, whereas the metalens-based marker provides a strong, well-localized response across all tested angles up to $15^\circ$.}
    \label{fig:imaging_results}
 %\vspace{-0.3cm}
\end{figure*}

To this end, we conducted an outdoor imaging experiment in which the bicycle was placed in front of a TI mmWave radar at a range of approximately $20~\mathrm{m}$, chosen to keep the bicycle within the radar's elevation $6~\mathrm{dB}$ beamwidth of $\pm 5^\circ$, as shown in Fig.~\ref{fig:setup_a}. In modern automotive systems, a radar is mounted on the vehicle's front bumper for collision avoidance. The height of the TI radar in our setup was selected to match this bumper height, as illustrated in Fig.~\ref{fig:setup_c}. The azimuthal angle of the bicycle was varied from $0^\circ$ (broadside to the radar) up to $15^\circ$ in steps of $5^\circ$, as this range captures the most commonly encountered relative orientations between vehicles and bicycles in realistic roadway scenarios. At each angle, two measurements were recorded: one with the metalens-based radar marker attached to the back seat of the bicycle, as shown in Fig.~\ref{fig:setup_b}, and one with the marker removed. A Hamming window was applied along the virtual-array antenna dimension to suppress spectral sidelobes, and the range--azimuth maps showing the linear-scale power of the 2D FFT were plotted. The images without and with the marker shown in Fig.~\ref{fig:a}--\ref{fig:d} and Fig.~\ref{fig:e}--\ref{fig:h}, respectively. As can be observed in Fig.~\ref{fig:imaging_results}, when the marker is absent (top row), the radar return from the bicycle alone is weak and barely distinguishable from the background clutter. In contrast, when the marker is attached (bottom row), a strong, well-localized return is clearly visible both at normal incidence and across all oblique angles tested. The marker enhances the peak 2D FFT power by approximately $20.44~\mathrm{dB}$ at $0^\circ$, $14.06~\mathrm{dB}$ at $5^\circ$, $9.03~\mathrm{dB}$ at $10^\circ$, and $6.02~\mathrm{dB}$ at $15^\circ$ relative to the strongest return from the bare bicycle in the same range window. These improvements translate into approximately $3.24\times$, $2.25\times$, $1.68\times$, and $1.41\times$ longer detection ranges at $0^\circ$, $5^\circ$, $10^\circ$, and $15^\circ$, respectively.

\vspace{0.3cm}

%\vspace{+0.1cm}

These results highlight the limited radar visibility of bare bicycles, particularly at broadside incidence where cyclists are commonly encountered by approaching vehicles, underscoring the importance of enhancing bicycle detectability for automotive radar safety applications. The proposed retrodirective metalens-based radar marker addresses this limitation by providing a 110x to 4x enhancement in backscatter response at broadside and up to $15^\circ$, offering a practical means to ensure detectability of cyclists in such scenarios.

\section*{Discussion and Conclusion}
\label{sec:conclusion}

In this paper, a compact, lightweight ($0.61~\mathrm{g}$) retrodirective safety reflector based on a single-layer phase-gradient metalens was demonstrated at $78.5~\mathrm{GHz}$. The proposed architecture, in which an incident plane wave is focused by the metalens onto a matching patch antenna pixel and re-radiated back through the same metalens, achieves stable retrodirective behavior in a thin, easily fabricated form factor that conventional mmWave reflectors have struggled to match. The peak RCS of $3.54~\mathrm{dBsm}$ and stable response over an $80^{\circ}$ angular range translate directly into long-range detectability up to approximately $72~\mathrm{m}$, and reliable imaging of objects at oblique angles that, without the tag, would otherwise be lost in clutter. Practically, the marker extends the detection range of a bare bicycle by up to $3.24\times$, demonstrating the large potential benefit of such markers to enhance the safety of cyclists among radar-assisted vehicles. As shown in Table~\ref{tab:comparison}, these properties place the proposed marker ahead of the state of the art, validating the metalens-based architecture as a viable building block for dedicated electromagnetic radar markers in next-generation automotive radar and wireless sensing systems. Future work will focus on incorporating modulation into the tag to enable unique signatures and identification, and on evaluating system performance under vehicular mobility, where the radar is mounted on a moving car.
\bibliography{sample}

@techreport{ntsb2019uber,
  author      = {{National Transportation Safety Board}},
  title       = {Collision Between Vehicle Controlled by Developmental Automated Driving System and Pedestrian, {Tempe, Arizona}, {March 18, 2018}},
  institution = {National Transportation Safety Board},
  year        = {2019},
  type        = {Highway Accident Report},
  number      = {NTSB/HAR-19/03},
  address     = {Washington, D.C.},
  url         = {https://www.ntsb.gov/investigations/accidentreports/reports/har1903.pdf}
}

@misc{reuters_waymo_2024,
  author       = {{Reuters}},
  title        = {Driverless {Waymo} car hits cyclist in {San Francisco}, causes minor scratches},
  year         = {2024},
  month        = {February},
  day          = {7},
  url          = {https://reuters.com},
  note         = {Accessed: May 17, 2026}
}

@article{feng2018lane,
  title={Lane detection with a high-resolution automotive radar by introducing a new type of road marking},
  author={Feng, Zhaofei and Li, Mingkang and Stolz, Martin and Kunert, Martin and Wiesbeck, Werner},
  journal={IEEE Transactions on Intelligent Transportation Systems},
  volume={20},
  number={7},
  pages={2430--2447},
  year={2018},
  publisher={IEEE}
}

@article{douglas2023low,
  title={A Low-Profile Passive Radar Reflector for Detection and Identification of Road Markings},
  author={Douglas, Tanner J and Sarabandi, Kamal},
  journal={IEEE Transactions on Antennas and Propagation},
  year={2023},
  publisher={IEEE}
}

@INPROCEEDINGS{IEEERFID2024,
  author={Ghasemi, Sepideh and Guo, Longyu and Harisha, Skanda and Eid, Aline},
  booktitle={2024 IEEE International Conference on RFID (RFID)}, 
  title={mmWave Retroreflective Road Markers for Automotive Radar Vision}, 
  year={2024},
  volume={},
  number={},
  pages={89-94},
  doi={10.1109/RFID62091.2024.10582666}}

@book{balanis2015antenna,
  title={Antenna theory: analysis and design},
  author={Balanis, Constantine A},
  year={2015},
  publisher={John wiley \& sons}
}

@misc{3M_Marker,
  note = {},
  author = {{3M™ Raised Pavement Marker Series 290}},
  year = {},
  howpublished = "{\url{https://www.3m.com/3M/en_US/p/d/b00010878/}}",
}

@article{arbabi2017planar,
  title={Planar metasurface retroreflector},
  author={Arbabi, Amir and Arbabi, Ehsan and Horie, Yu and Kamali, Seyedeh Mahsa and Faraon, Andrei},
  journal={Nature Photonics},
  volume={11},
  number={7},
  pages={415--420},
  year={2017},
  publisher={Nature Publishing Group UK London}
}

@article{lynch2024intersection,
  title={At the Intersection Between Optics and mmWave Design: An Energy Autonomous 5G-Enabled Multilens-Based Broadbeam mmID for “Smart” Digital Twins Applications},
  author={Lynch, Charles A and Soto-Valle, Genaro and Hester, Jimmy GD and Tentzeris, Manos M},
  journal={IEEE Transactions on Microwave Theory and Techniques},
  volume={72},
  number={4},
  pages={2620--2630},
  year={2024},
  publisher={IEEE}
}

@article{yu2011light,
  title={Light propagation with phase discontinuities: generalized laws of reflection and refraction},
  author={Yu, Nanfang and Genevet, Patrice and Kats, Mikhail A and Aieta, Francesco and Tetienne, Jean-Philippe and Capasso, Federico and Gaburro, Zeno},
  journal={science},
  volume={334},
  number={6054},
  pages={333--337},
  year={2011},
  publisher={American Association for the Advancement of Science}
}

@article{chen2024focus,
  title={Focus control of wide-angle metalens based on digitally encoded metasurface},
  author={Chen, Yi and Zhang, Simeng and Tian, Ying and Li, Chenxia and Huang, Wenlong and Liu, Yixin and Jin, Yongxing and Fang, Bo and Hong, Zhi and Jing, Xufeng},
  journal={Opto-Electronic Advances},
  volume={7},
  number={8},
  pages={240095--1},
  year={2024},
  publisher={Opto-Electronic Advances}
}

@article{lee2021single,
  title={Single-layer phase gradient mmWave metasurface for incident angle independent focusing},
  author={Lee, Wonwoo and Jo, Semin and Lee, Kanghyeok and Park, Hong Soo and Yang, Junhyuk and Hong, Ha Young and Park, Changkun and Hong, Sun K and Lee, Hojin},
  journal={Scientific reports},
  volume={11},
  number={1},
  pages={12671},
  year={2021},
  publisher={Nature Publishing Group UK London}
}

@article{cai2017high,
  title={High-efficiency and full-space manipulation of electromagnetic wave fronts with metasurfaces},
  author={Cai, Tong and Wang, GuangMing and Tang, ShiWei and Xu, HeXiu and Duan, JingWen and Guo, HuiJie and Guan, FuXin and Sun, ShuLin and He, Qiong and Zhou, Lei},
  journal={Physical Review Applied},
  volume={8},
  number={3},
  pages={034033},
  year={2017},
  publisher={APS}
}

@article{cheng2019realizing,
  title={Realizing broadband transparency via manipulating the hybrid coupling modes in metasurfaces for high-efficiency metalens},
  author={Cheng, Kaiyang and Wei, Zeyong and Fan, Yuancheng and Zhang, Xiaoming and Wu, Chao and Li, Hongqiang},
  journal={Advanced Optical Materials},
  volume={7},
  number={15},
  pages={1900016},
  year={2019},
  publisher={Wiley Online Library}
}

@ARTICLE{JRFID,
  author={Ghasemi, Sepideh and Guo, Longyu and Hester, Jimmy G. D. and Eid, Aline},
  journal={IEEE Journal of Radio Frequency Identification}, 
  title={Retrodirective Electromagnetic Pavement Markers for Enhanced Automotive Radar Vision}, 
  year={2025},
  volume={9},
  number={},
  pages={395-406},
  doi={10.1109/JRFID.2025.3575348}}

@article{Grbic,
  title = {Metamaterial Huygens' Surfaces: Tailoring Wave Fronts with Reflectionless Sheets},
  author = {Pfeiffer, Carl and Grbic, Anthony},
  journal = {Phys. Rev. Lett.},
  volume = {110},
  issue = {19},
  pages = {197401},
  numpages = {5},
  year = {2013},
  month = {May},
  publisher = {American Physical Society},
  doi = {10.1103/PhysRevLett.110.197401},
  url = {https://link.aps.org/doi/10.1103/PhysRevLett.110.197401}
}

@article{pfeiffer2013millimeter,
  title   = {Millimeter-wave transmitarrays for wavefront and polarization control},
author = {Pfeiffer, Carl and Grbic, Anthony},
journal = {IEEE Transactions on Microwave Theory and Techniques},
volume  = {61}, number = {12}, pages = {4407--4417},
year    = {2013}}

@article{li2020characterization,
  title={Characterization of metasurface lens antenna for sub-6 GHz dual-polarization full-dimension massive MIMO and multibeam systems},
  author={Li, Shunli and Chen, Zhi Ning and Li, Teng and Lin, Feng Han and Yin, Xiaoxing},
  journal={IEEE Transactions on Antennas and Propagation},
  volume={68},
  number={3},
  pages={1366--1377},
  year={2020},
  publisher={IEEE}
}

@article{pesarakloo2022planar,
  title={Planar, wide-band omnidirectional retroreflector using metal-only transmitarray structure for TE and TM polarizations},
  author={Pesarakloo, Ali and Khalaj-Amirhosseini, Mohammad},
  journal={Scientific Reports},
  volume={12},
  number={1},
  pages={11279},
  year={2022},
  publisher={Nature Publishing Group UK London}
}

@article{liu2020design,
  title={Design and demonstration of a wide-angle and high-efficient planar metasurface lens},
  author={Liu, Yong-Qiang and Che, Yongxing and Qi, Kainan and Li, Liangsheng and Yin, Hongcheng},
  journal={Optics Communications},
  volume={474},
  pages={126061},
  year={2020},
  publisher={Elsevier}
}

@article{liu2022low,
  title={Low-profile and compact retroreflector enabled by a wide-angle and high-efficiency metalens},
  author={Liu, Yong-Qiang and Guo, Jie and Li, Sheng and Qi, Kainan and Li, Liangsheng and Yin, Hongcheng},
  journal={Optical Materials},
  volume={134},
  pages={113105},
  year={2022},
  publisher={Elsevier}
}

@article{datta2022gradient,
  title={Gradient index metasurface lens for microwave imaging},
  author={Datta, Srijan and Tamburrino, Antonello and Udpa, Lalita},
  journal={Sensors},
  volume={22},
  number={21},
  pages={8319},
  year={2022},
  publisher={MDPI}
}

@article{guo2018high,
  title={High-efficiency and wide-angle beam steering based on catenary optical fields in ultrathin metalens},
  author={Guo, Yinghui and Ma, Xiaoliang and Pu, Mingbo and Li, Xiong and Zhao, Zeyu and Luo, Xiangang},
  journal={Advanced Optical Materials},
  volume={6},
  number={19},
  pages={1800592},
  year={2018},
  publisher={Wiley Online Library}
}

@article{Khorasaninejad2017,
  title = {Metalenses: Versatile multifunctional photonic components},
  author = {Khorasaninejad, Mohammad and Capasso, Federico},
  journal = {Science},
  volume = {358},
  number = {6367},
  pages = {eaam8100},
  year = {2017},
  publisher = {American Association for the Advancement of Science}
}

@article{hallbjorner2013,
  author={Hallbj\"orner, P. and Cheng, S.},
  journal={IEEE Antennas and Wireless Propagation Letters},
  title={Improvement in 77-GHz Radar Cross Section of Road Work Jacket and Side Screen by Use of Planar Flexible Retrodirective Reflectors},
  year={2013}, volume={12}, pages={1085--1088}
}

@article{braun2023harmonic,
  author={Braun, T. T. and Sch\"opfel, J. and Kwiatkowski, P. and Schweer, C. and Aufinger, K. and Pohl, N.},
  journal={IEEE Transactions on Microwave Theory and Techniques},
  title={Expanding the Capabilities of Automotive Radar for Bicycle Detection With Harmonic RFID Tags at 79/158 GHz},
  year={2023}, volume={71}, number={1}, pages={320--331}
}

@misc{vovchuk2026optically,
  author={Vovchuk, D. and others},
  title={Optically Transparent Meta-Grating Embedded in Rear Windshields for Automotive Radar Detection},
  year={2026},
  eprint={2601.01551},
  archivePrefix={arXiv},
  url={https://arxiv.org/abs/2601.01551}
}

@misc{ti_awr2944evm,
  author={{Texas Instruments}},
  title={{AWR2944EVM} Evaluation Module: $76$--$81~\mathrm{GHz}$ Automotive Radar Sensor with $4$ {Tx} and $4$ {Rx} Channels},
  howpublished={\url{https://www.ti.com/tool/AWR2944EVM}},
  year={2023},
  note={Accessed: May 2026}
}

@article{yang2025design,
  author={Yang, Xuan and Yan, Jinfeng and Fan, Zepeng and Wang, Yifan and Lv, Songtao and Ong, Ghim Ping and Wang, Dawei},
  journal={IEEE Transactions on Intelligent Transportation Systems},
  title={Design of Electromagnetic Road Markings for Implementing 77 {GHz} Millimeter-Wave Radar Sensing},
  year={2025},
  volume={26},
  number={10},
  pages={17150--17160},
  doi={10.1109/TITS.2025.3578049}
}

%\noindent LaTeX formats citations and references automatically using the bibliography records in your .bib file, which you can edit via the project menu. Use the cite command for an inline citation, e.g.  \cite{Hao:gidmaps:2014}.

%For data citations of datasets uploaded to e.g. \emph{figshare}, please use the \verb|howpublished| option in the bib entry to specify the platform and the link, as in the \verb|Hao:gidmaps:2014| example in the sample bibliography file.

%\section*{Author contributions statement}

%Must include all authors, identified by initials, for example:
%A.A. conceived the experiment(s),  A.A. and B.A. conducted the experiment(s), C.A. and D.A. analysed the results.  All authors reviewed the manuscript. 

\end{document}